\documentclass[fleqn,usenatbib]{mnras}

\usepackage{newtxtext,newtxmath}
\usepackage[T1]{fontenc}

\DeclareRobustCommand{\VAN}[3]{#2}
\let\VANthebibliography\thebibliography
\def\thebibliography{\DeclareRobustCommand{\VAN}[3]{##3}\VANthebibliography}

\usepackage{graphicx}	% Including figure files
\usepackage{amsmath}	% Advanced maths commands
\usepackage{balance} % For \balance command

\title[Gas mixing through a SPH approach]{Gas mixing through a Smoothed Particle Hydrodynamics approach}

\author[L. Maggioni et al.]{
L. Maggioni,$^{1,2}$\thanks{E-mail:luca.maggioni@inaf.it}
M. Teodori,$^{1,3}$
G. Magni,$^{1}$
M. Formisano,$^{1}$
M.C. De Sanctis$^{1}$
and F. Altieri$^{1}$
\\
$^{1}$INAF-IAPS, Via del Fosso del Cavaliere 100, 00133, Rome, Italy\\
$^{2}$Department of Physics, University of Rome Tor Vergata, Via della Ricerca Scientifica 1, 00133, Rome, Italy\\
$^{3}$Department of Mathematics and Physics, University of Campania Luigi Vanvitelli, Viale Lincoln 5, 81100, Caserta, Italy 
}
\date{Accepted 2025 August 05. Received 2025 July 29; in original form 2025 May 09}

\pubyear{\the\year{}}

\begin{document}
\label{firstpage}
\pagerange{\pageref{firstpage}--\pageref{lastpage}}
\maketitle

\begin{abstract}
Transport and mixing of gas species are of particular interest in planetary environments, where interactions among multiple species can occur within confined or porous media. In this work, we present a novel Smoothed Particle Hydrodynamics (SPH) approach for modeling the mixing of binary gas species. The model treats each gas as a separate fluid governed by its own set of Euler equations, coupled through collisional momentum and energy exchange terms derived from a kinetic relaxation model based on the Boltzmann equation.
The numerical scheme employs a first-order operator splitting approach combined with a two-step Euler integrator. In this setup, the hydrodynamic evolution is first computed using standard SPH techniques to handle pressure forces. This is followed by a separate correction step that accounts for interspecies collisional exchanges. Such a decoupled treatment enables the use of a larger timestep dictated by hydrodynamics rather than the typically much smaller collisional timescale, enhancing computational efficiency.
The model achieves good accuracy in reproducing the equilibration of density and temperature in a range of molecular mass ratios. Its modular structure supports natural extensions to polyatomic mixtures and enables the inclusion of additional physics, such as gas-solid interactions with dust and ice. These features make the method particularly well-suited for applications involving confined, multi-component gas systems, such as those expected during the ESA ExoMars mission.
\end{abstract}

\begin{keywords}
Hydrodynamics -- Numerical methods -- Planetary Systems
\end{keywords}

%%%%%%%%%%%%%%%%%%%%%%%%%%%%%%%%%%%%%%%%%%%%%%%%%%

%%%%%%%%%%%%%%%%% BODY OF PAPER %%%%%%%%%%%%%%%%%%

\section{Introduction}

Many fluid dynamics problems, especially in planetary science, involve multiple phases and components, and require robust numerical methods. Lagrangian methods like Smoothed Particle Hydrodynamics (SPH) \citep{GingoldMonaghan1977, Lucy1977, Monaghan2005} offer great flexibility, making them well-suited for problems involving multiple components. In the SPH framework, the fluid is discretized into a set of pseudo-particles, and the temporal evolution of physical properties is tracked by following these particles. The properties of each particle are determined from those of all neighboring particles lying within the range controlled by an interpolation kernel function.\\
Several SPH-based studies have addressed mixing phenomena in multiphase and multi-component systems. These include applications to solid–liquid, gas–liquid, and multi-fluid interactions \citep{Parkash2007, Ren2014, Ai2018, Bertevas2019}, as well as models that represent mixing by tracking the concentration of each component \citep{ Greif2009, Kwon2019}. Efforts have also been made to improve numerical treatment of mixing across discontinuities or unstable interfaces, often in the context of turbulent flows \citep{Read2010, SANDNES2025}.
Another important class of SPH models addresses gas–dust interactions \citep{MONAGHAN1995, LP2012a, LP2012b}, where drag forces govern the momentum and energy coupling between phases. While \citet{MONAGHAN1995} noted that such models could be extended to gas–gas mixtures, doing so introduces further complexity: the timescale of molecular collisions is typically much smaller than the characteristic hydrodynamic timescales. This poses a challenge when modeling the diffusion of two gases in a broader context that includes additional phases and other physical processes.\\
In the present work, we build upon the structure of gas–dust models but extend it to a new physical regime. We introduce an SPH-based model specifically designed to simulate binary gas mixtures, where collisional momentum and energy exchange terms are derived from a kinetic relaxation model based on the Boltzmann equation. In particular, the model employs a multi-fluid formulation, with separate sets of equations for each gas species. This method enables a detailed representation of the flow fields for each constituent individually, making it particularly advantageous in non-equilibrium conditions where the flow characteristics of the gases can differ significantly. Additionally, to efficiently handle the much shorter diffusion timescale, we implement a splitting scheme, allowing the system to evolve using the hydrodynamic timestep while significantly reducing the computational cost of the simulation.\\
The approach presented here is applicable to any binary mixture of monatomic gases and is intended as a first step toward a more comprehensive model that accounts for mixing between polyatomic gases, which requires the inclusion of additional contributions from rotational and vibrational degrees of freedom. The mixing of such gases plays a significant role in a wide range of astrophysical and planetary contexts -- for example, in the interaction between volatiles (e.g., water vapor, methane) and the Mars' $\mathrm{CO_2}$-dominated atmosphere.

\section{Model}
\label{subsec:atmospheric effects}

Before entering into the details of the computational model, we first provide an explanation of the theoretical framework employed in this work.\\
Consider a system composed of two gases, denoted as $\mathrm{\alpha}$ and $\mathrm{\beta}$ for simplicity. Since the gases are treated as ideal, intermolecular collisions are assumed to be negligible by definition. However, if molecular collisions were entirely absent, the molecules of each species in an ideal gas mixture would be insensible to the presence of the other species. Therefore, to enhance the realism of the model, we include the effects of collisions between molecules of different species, as proposed in \citet{Ramshaw2002}.\\ 
In particular, we assume that molecular interactions are governed by a Lennard-Jones (12,6) potential \citep{Chapman1970}:
\begin{equation}
\label{LennardJones}
    \psi(r) = \epsilon_{\alpha\beta}\left[\left(\frac{\sigma_{\alpha\beta}}{r}\right)^{12}-\left(\frac{\sigma_{\alpha\beta}}{r}\right)^{6}\right],
\end{equation}
where $r$ is the intermolecular separation, $\sigma_{\alpha\beta}$ is the distance at which the particle-particle potential energy becomes zero, and $\epsilon_{\alpha\beta}$ represents the maximum energy of attraction. The parameters $\epsilon$ and $\sigma$ are intrinsic properties of each gas. Nevertheless, for a mixture of two gases, the corresponding $\epsilon_{\alpha\beta}$ and $\sigma_{\alpha\beta}$ in Eq. \eqref{LennardJones} are given by
\begin{align}
    \epsilon_{\alpha\beta} = \sqrt{\epsilon_{\alpha}\epsilon_{\beta}} \; , \quad
    \sigma_{\alpha\beta} = \frac{\sigma_{\alpha} + \sigma_{\beta}}{2} 
\label{eps&sigmaMix}.
\end{align}
The model presented in this work is a two-fluid model, in which the density, momentum, and energy of each gas are treated separately. While the equations governing the density remain unchanged, the standard Euler equations for a pure, inviscid ideal gas with zero thermal conductivity are modified to include collisional terms that account for momentum and energy exchange between the two gases:
\begin{align}
\label{gas_gas_system}
\begin{split}
     \frac{\mathrm{d}\rho_{\alpha}}{\mathrm{d}t}
     = &
     - \rho_{\alpha} \left(\nabla\cdot\mathbf{v}_{\alpha}\right),\\[10pt]
     \frac{\mathrm{d}\rho_{\beta}}{\mathrm{d}t}
     = &
     - \rho_{\beta} \left(\nabla\cdot\mathbf{v}_{\beta}\right),\\[10pt]
     \frac{\mathrm{d}\mathbf{v}_{\alpha}}{\mathrm{d}t}
     = & 
     -\frac{\nabla P_{\alpha}}{\rho_{\alpha}} +  \left(\frac{\mathrm{d}\mathbf{v}_{\alpha}}{\mathrm{d}t}\right)_{\mathrm{coll}}\\[10pt]
     \frac{\mathrm{d}\mathbf{v}_{\beta}}{\mathrm{d}t}
     = & 
     -\frac{\nabla P_{\beta}}{\rho_{\beta}} +  \left(\frac{\mathrm{d}\mathbf{v}_{\beta}}{\mathrm{d}t}\right)_{\mathrm{coll}}, \\[10pt]
     \frac{\mathrm{d}\epsilon_{\mathrm{tk},\alpha}}{\mathrm{d}t}
     = & 
     -\frac{1}{\rho_{\alpha}}\nabla\cdot\left( P_{\alpha}\mathbf{v}_{\alpha}\right)+  \left(\frac{\mathrm{d}{\epsilon_{\mathrm{tk},\alpha}}}{\mathrm{d}t}\right)_{\mathrm{coll}},\\[10pt]
     \frac{\mathrm{d}\epsilon_{\mathrm{tk},\beta}}{\mathrm{d}t}
     = & 
     -\frac{1}{\rho_{\beta}}\nabla\cdot\left( P_{\beta}\mathbf{v}_{\beta}\right)+  \left(\frac{\mathrm{d}{\epsilon_{\mathrm{tk},\beta}}}{\mathrm{d}t}\right)_{\mathrm{coll}}.
\end{split}
\end{align}
To close the system, the ideal gas equation of state (EOS) is applied to each gas:
\begin{equation}
P_{i} = c_{\mathrm{s},i}^2 \frac{\rho_{i}}{\gamma_{i}}  \qquad i = \alpha, \beta.
\label{eq:EoSgas}
\end{equation}
In these expressions, $\rho$ represents the density, $\mathbf{v}$ the velocity, $P$ the pressure, and $\epsilon_{\mathrm{tk}}$ the thermokinetic energy per unit mass, which includes both kinetic and internal thermal energy ($\epsilon_{\mathrm{tk}} = \epsilon_{\mathrm{kin}} + \epsilon_{\mathrm{int}}$). The latter is related to the temperature by the relation $\epsilon_{\mathrm{int}} = c_{\mathrm{v}}T$, in which $c_{\mathrm{v}}$ is the specific heat at constant volume and $T$ is the temperature. The sound speed of each particle, instead, is determined by the relation $c_\mathrm{s} = \sqrt{\gamma P/\rho}=\sqrt{\gamma k_\mathrm{B}T/m}$, where $\gamma$ is the adiabatic index, $k_\mathrm{B}$ is the Boltzmann constant, and $m$ is the molecular mass.

\subsection{Mixing terms}

In gas kinetic theory, the evolution of a binary mixture is described by two velocity distribution functions $g(\mathbf{x}, \mathbf{u}, t)$, where $\mathbf{x}$ is the position, $\mathbf{u}$ the molecular velocity, and $t$ is the time. These distribution functions satisfy the collisional Boltzmann equation \citep{Chapman1970}, which includes terms for both self-collisions and cross-collisions. As assumed at the beginning of this discussion, self-collisions are neglected. Cross-collisions, on the other hand, are challenging to handle due to the complexity of the collision integral in the Boltzmann equation. In principle, this integral includes contributions from both elastic and inelastic interactions \citep{Reyes2007}. However, we restrict our analysis to the elastic component, which is appropriate given that we focus exclusively on monatomic gases, for which inelastic effects are negligible. To simplify the mathematical treatment, we adopt the kinetic relaxation model proposed by Gross and Krook \citep{Gross1956}, in which the elastic collision integral is approximated by a relaxation term of the form $f_{\alpha\beta}\left(g_{i = \alpha,\beta} - \hat{g}_{i = \alpha,\beta}\right)$. Here, $f_{\alpha\beta}$ is a velocity-independent effective collision frequency, and $\hat{g}$ represents an equilibrium distribution function. As noted in \citet{Reyes2007}, solving this kinetic model requires selecting a suitable form for $\hat{g}$. The most natural choice is to assume a Gaussian distribution:
\begin{equation}
    \hat{g} = n \left(\frac{m}{2\pi k_{B}\hat{T}}\right)^{3/2} \mathrm{exp}\left(-\frac{m}{2 k_{B} \hat{T}}\left(\mathbf{u}-\hat{\mathbf{v}}\right)^{2}\right),
\end{equation}
where $n$ is the number density, while $\hat{T}$ and $\hat{\mathbf{v}}$ are two collisional parameters. As demonstrated in \citet{Reyes2007, ZAHMATKESH2013}, ensuring that the proposed kinetic model correctly reproduces the collisional momentum and energy equations of the original elastic Boltzmann operator leads to the determination of $\hat{T}$ and $\hat{\mathbf{v}}$ for both gases $\alpha$ and $\beta$:
\begin{align}
\label{collisional_parameters}
\begin{split}
    \hat{\mathbf{v}}_{\alpha} = &  \hat{\mathbf{v}}_{\beta} =  \frac{m_{\alpha}\mathbf{v}_{\alpha}+m_{\beta}\mathbf{v}_{\beta}}{m_{\alpha}+m_{\beta}},\\[10pt]
    \hat{T}_{\alpha} = & T_{\alpha} + 2 \frac{m_{\alpha} m_{\beta}}{(m_{\alpha} + m_{\beta})^2} \left( (T_{\beta} - T_{\alpha}) + \frac{m_{\beta}}{6k_{B}} (\mathbf{v}_{\alpha} - \mathbf{v}_{\beta})^2 \right),\\[10pt]
    \hat{T}_{\beta} = & T_{\beta} + 2 \frac{m_{\alpha} m_{\beta}}{(m_{\alpha} + m_{\beta})^2} \left( (T_{\alpha} - T_{\beta}) + \frac{m_{\alpha}}{6k_{B}} (\mathbf{v}_{\beta} - \mathbf{v}_{\alpha})^2 \right).
\end{split}
\end{align}
The determination of these parameters completes the construction of the kinetic model.
At this stage, by taking moments of the Boltzmann equation -- multiplying it once by $1$, once by $\mathbf{u}$, once by $\mathbf{u}^{2}/2$, and integrating over momentum space --  the macroscopic conservation equations for mass, momentum, and energy (i.e., the Euler equations) can be derived. We refer the reader to \citet{Chapman1970} for the detailed calculations.

\subsection{Momentum exchange}

The collisional momentum exchange terms, derived from the Boltzmann equation, take the following form \citep{Chapman1970,ZAHMATKESH2013, Wargnier2022}:
\begin{align}
\label{coll_part}
\begin{split}
     \left(\frac{\mathrm{d}\mathbf{v}_{\alpha}}{\mathrm{dt}}\right)_{\mathrm{coll}}
     = & f_{\alpha\beta}\left(\hat{\mathbf{v}}_{\alpha} - \mathbf{v}_{\alpha}\right),
     \\[10pt]
     \left(\frac{\mathrm{d}\mathbf{v}_{\beta}}{\mathrm{dt}}\right)_{\mathrm{coll}}
     = & f_{\beta\alpha}\left(\hat{\mathbf{v}}_{\beta} - \mathbf{v}_{\beta}\right).
\end{split}
\end{align}
Here $f_{\alpha\beta}$ and $f_{\beta\alpha}$ are the frequencies of cross-collision between $\alpha$ and $\beta$, and are calculated as
\begin{align}
\label{int_term}
    \begin{split}
        f_{\alpha\beta} = & \frac{16}{3}\frac{\rho_{\beta}}{m_{\beta}}\Omega_{\alpha\beta}^{(1,1)} ,\\[10pt]
        f_{\beta\alpha} = & \frac{16}{3}\frac{\rho_{\alpha}}{m_{\alpha}}\Omega_{\beta\alpha}^{(1,1)},
    \end{split}
\end{align} 
where $\Omega_{\alpha\beta}^{(i,j)}$ denote the Chapman-Cowling collision integrals, which, under the assumption of a Lennard-Jones intermolecular potential, are given by \citep{Chapman1970}
\begin{equation}
\label{chapman_cowling_int}
    \Omega_{\alpha\beta}^{(i,j)} = \sigma_{\alpha\beta}^{2} \left(\frac{2\pi k_{B}T_{\mathrm{mix}}}{m_{\alpha\beta}}\right)^{1/2} W_{\alpha\beta}^{(i,j)}(T^{*}).
\end{equation}
Here, $m_{\alpha\beta} = m_{\alpha}m_{\beta}/(m_{\alpha}+m_{\beta})$ is the reduced molecular mass, $T_{\mathrm{mix}} = (n_{\alpha}T_{\alpha} + n_{\beta}T_{\beta}) / (n_{\alpha} + n_{\beta})$ is the mixture temperature, and $W_{\alpha\beta}^{(i,j)}(T^{*})$ is a non-dimensional integral, which depends on the reduced temperature $T^{*} = k_{B}T_{\mathrm{mix}}/\epsilon_{\alpha\beta}$. Empirical expressions for computing these integrals are provided by \citet{Neufield1972}. In particular, since only the integral with $(i=1,j=1)$ is required in Eq. \eqref{int_term}, the shape of $W_{\alpha\beta}^{(1,1)}(T^{*})$ is:
\begin{equation}
\label{tabulated_W}
    W_{\alpha\beta}^{(1,1)}(T^{*}) = \frac{A}{(T^{*})^{B}} + \frac{C}{\exp(D\;T^{*})} + \frac{E}{\exp(F\;T^{*})} + \frac{G}{\exp(H\;T^{*})},
\end{equation}
with $A = 1.06036$, $B = 0.15610$, $C = 0.19300$, $D = 0.47635$, $E = 1.03587$, $F = 1.52996$, $G = 1.76474$, $H = 3.89411$.\\
Putting together Eqs. \eqref{int_term}, \eqref{chapman_cowling_int}, \eqref{tabulated_W}, and using the definition of $\hat{\mathbf{v}}$, the momentum equations in Eq. \eqref{gas_gas_system} become
\begin{align}
\label{pressure_coll_system}
\begin{split}
     \frac{\mathrm{d}\mathbf{v}_{\alpha}}{\mathrm{d}t}
     = & 
     -\frac{\nabla P_{\alpha}}{\rho_{\alpha}} - \frac{16 \;\sigma_{\alpha\beta}^{2}\left(2\pi k_{B}\right)^{1/2}}{3 \sqrt{m_{\alpha}m_{\beta}(m_{\alpha}+m_{\beta})}} \rho_{\beta}T_{\mathrm{mix}}^{1/2} W_{\alpha\beta}^{(1,1)}(\mathbf{v_{\alpha}-\mathbf{v_{\beta}}}) ,\\[10pt]
     \frac{\mathrm{d}\mathbf{v}_{\beta}}{\mathrm{d}t}
     = & 
     -\frac{\nabla P_{\beta}}{\rho_{\beta}} + \frac{16 \;\sigma_{\alpha\beta}^{2}\left(2\pi k_{B}\right)^{1/2}}{3 \sqrt{m_{\alpha}m_{\beta}(m_{\alpha}+m_{\beta})}} \rho_{\alpha}T_{\mathrm{mix}}^{1/2} W_{\beta\alpha}^{(1,1)}(\mathbf{v_{\alpha}-\mathbf{v_{\beta}}}) .
\end{split}
\end{align}
Nevertheless, since $W_{\alpha\beta}^{(1,1)} = W_{\beta\alpha}^{(1,1)}$, we can define a mixture coefficient
\begin{equation}
\label{mixture_coefficient}
    K_{\mathrm{mix}} = \frac{16 \;\sigma_{\alpha\beta}^{2}\left(2\pi k_{B}\right)^{1/2}}{3 \sqrt{m_{\alpha}m_{\beta}(m_{\alpha}+m_{\beta})}} T_{\mathrm{mix}}^{1/2} W_{\alpha\beta}^{(1,1)},
\end{equation}
such that the Eqs. \eqref{pressure_coll_system} can be easier written as
\begin{align}
\label{pressure_coll_system_simple_form}
\begin{split}
     \frac{\mathrm{d}\mathbf{v}_{\alpha}}{\mathrm{d}t}
     = & 
     -\frac{\nabla P_{\alpha}}{\rho_{\alpha}} - \rho_{\beta}K_{\mathrm{mix}}(\mathbf{v_{\alpha}-\mathbf{v_{\beta}}}) ,\\[10pt]
     \frac{\mathrm{d}\mathbf{v}_{\beta}}{\mathrm{d}t}
     = & 
     -\frac{\nabla P_{\beta}}{\rho_{\beta}} + \rho_{\alpha}K_{\mathrm{mix}}(\mathbf{v_{\alpha}-\mathbf{v_{\beta}}}).
\end{split}
\end{align}
The symmetry between the two collisional terms in these equations is evident and guarantees the conservation of total momentum:
\begin{equation}
    \rho_{\alpha}\frac{\mathrm{d}\mathbf{v}_{\alpha}}{\mathrm{d}t} + \rho_{\beta}\frac{\mathrm{d}\mathbf{v}_{\beta}}{\mathrm{d}t} = - \nabla\left(P_{\alpha} + P_{\beta}\right).
\end{equation}

\subsection{Energy exchange}

Concerning energy, the collisional exchange terms are given by \citep{ZAHMATKESH2013}:
\begin{align}
\begin{split}
\label{energy_analytical}
       \left(\frac{\mathrm{d}{\epsilon_{\mathrm{tk},\alpha}}}{\mathrm{d}t}\right)_{\mathrm{coll}} &= f_{\alpha\beta} \left(\hat{\epsilon}_{\alpha} - \epsilon_{\mathrm{tk},\alpha}\right), \\[10pt]
     \left(\frac{\mathrm{d}{\epsilon_{\mathrm{tk},\beta}}}{\mathrm{d}t}\right)_{\mathrm{coll}} &= f_{\beta\alpha} \left(\hat{\epsilon}_{\beta} - \epsilon_{\mathrm{tk},\beta}\right),
\end{split}    
\end{align}
where $\hat{\epsilon}$ are defined as
\begin{align}
\label{epsilon_hat}
\begin{split}
    \hat{\epsilon}_{\alpha} = c_{\mathrm{v}, \alpha}\hat{T}_{\alpha} + \frac{\hat{v}_{\alpha}^{2}}{2}, \\[10pt]
    \hat{\epsilon}_{\beta} = c_{\mathrm{v}, \beta}\hat{T}_{\beta} + \frac{\hat{v}_{\beta}^{2}}{2}.
\end{split}
\end{align}
In contrast to the formulation in \citet{ZAHMATKESH2013}, the collisional terms $\hat{\epsilon}$ are defined here as energy per unit mass rather than per unit volume.
Although the symmetry between the Eqs. \eqref{energy_analytical} is less explicit than in the momentum exchange terms reformulated in Eq. \eqref{pressure_coll_system_simple_form}, it can be shown -- by employing the relevant parameter definitions -- that the following equality holds:
\begin{equation}
    \rho_{\alpha}f_{\alpha\beta}\left(\hat{\epsilon}_{\alpha} - \epsilon_{\mathrm{tk},\alpha}\right) + \rho_{\beta}f_{\beta\alpha} \left(\hat{\epsilon}_{\beta} - \epsilon_{\mathrm{tk},\beta}\right) = 0,
\end{equation}
which ensures the conservation of total energy:
\begin{equation}
    \rho_{\alpha}\frac{\mathrm{d}\epsilon_{\mathrm{tk},\alpha}}{\mathrm{d}t} + \rho_{\beta}\frac{\mathrm{d}\epsilon_{\mathrm{tk},\beta}}{\mathrm{d}t} = -\nabla\cdot\left( P_{\alpha}\mathbf{v}_{\alpha} + P_{\beta}\mathbf{v}_{\beta}\right).
\end{equation}

\subsection{Collisional timestep}
\label{collisional_timestep}
Once the Euler equations include the collisional exchange terms, we must determine the new constraint imposed on the integration timestep. To achieve this, we could follow an empirical approach, as suggested by \citet{MONAGHAN1995}, but we prefer a more formal derivation, such as the one used for the drag force by \citet{LP2012a}. Indeed, in a similar way, we analyze the stability of a simple forward Euler explicit scheme using a standard Von Neumann analysis.\\
The time discretization of the momentum collisional part from step $n$ to $n+1$ is given by
\begin{align}
\label{time_discretization_stability}
\begin{split}
     \frac{\mathbf{v}_{\alpha}^{n+1}-\mathbf{v}_{\alpha}^{n}}{\mathrm{\Delta}t}
     = & 
     - \rho_{\beta}K_{\mathrm{mix}}(\mathbf{v}_{\alpha}^{n}-\mathbf{v}_{\beta}^{n}) ,\\[10pt]
      \frac{\mathbf{v}_{\beta}^{n+1}-\mathbf{v}_{\beta}^{n}}{\mathrm{\Delta}t}
     = & 
     + \rho_{\alpha}K_{\mathrm{mix}}(\mathbf{v}_{\alpha}^{n}-\mathbf{v}_{\beta}^{n}).
\end{split}
\end{align}
We then introduce a perturbation to the velocity field relative to equilibrium at timestep $m$, assuming a monochromatic plane wave. Specifically, we have:
\begin{align}
\label{monocromatic_waves}
    \begin{split}
        \mathbf{v}_{\alpha}^{m} =& \mathbf{V}_{\alpha}^{m} e^{i k x}, \\
        \mathbf{v}_{\beta}^{m} = & \mathbf{V}_{\beta}^{m} e^{i k x},
    \end{split}
\end{align}
where, $\mathbf{V}_{\alpha}$ and $\mathbf{V}_{\beta}$ are complex constants, and $k$ is the wavenumber. Substituting these relations into Eq. \eqref{time_discretization_stability} results in a linear system with a matrix $\mathcal{M}$
\begin{equation}
\label{matrix_eigenvalues}
\mathcal{M} = 
    \begin{pmatrix}
        1-\rho_{\beta} K_{\mathrm{mix}}\mathrm{\Delta}t  & \rho_{\beta} K_{\mathrm{mix}}\mathrm{\Delta}t \\
        \rho_{\alpha} K_{\mathrm{mix}}\mathrm{\Delta}t & 1- \rho_{\alpha} K_{\mathrm{mix}}\mathrm{\Delta}t
    \end{pmatrix}.
\end{equation}
Stability is guaranteed when the smallest eigenvalue of $\mathcal{M}$ is smaller than $1$ in module. This condition translates in the following constraint for the collisional timestep:
\begin{equation}
\label{collitional_timestep}
     \mathrm{d}t_{\mathrm{coll}} < \frac{2}{K_{\mathrm{mix}}(\rho_{\alpha}+\rho_{\beta})} .
\end{equation}

\subsection{Numerical implementation}
\label{Numerical_implementation}

The SPH formalism is well established in the literature, and readers interested in a detailed explanation and derivation of the standard SPH form of Euler equations are referred to existing works \citep[e.g.,][]{Monaghan2005}. We now turn our attention to formulating the collisional terms within the Lagrangian formalism, as this has not yet been addressed in the literature.\\
Looking at Eq.\eqref{pressure_coll_system_simple_form}, we can observe that the structure of the collisional momentum exchange terms is formally identical to that of the drag acceleration proposed in \citet{MONAGHAN1995, LP2012a}. In these works, drag refers to the dynamical interaction between a solid component (e.g., dust) and a gas component. In such cases, the momentum exchange terms in the Euler equations are proportional to $K_{\mathrm{drag}}\Delta\mathbf{v}$, matching the form of the collisional terms in our model. This formal analogy allows us, in principle, to adopt a similar approach to the drag implementation described by \citet{MONAGHAN1995}. Following the convention where the indices $\{a,b\}$ refer to quantities computed for particles of gas $\alpha$, while $\{i,j\}$ refer to quantities computed for $\beta$ gas particles, the SPH formulation of the collisional acceleration part in Eq. \ref{pressure_coll_system_simple_form} is given by
\begin{align}
\label{collisional sph}
\begin{split}
     \left(\frac{\mathrm{d}\mathbf{v}_{\alpha}}{\mathrm{d}t}\right)_{\mathrm{coll}} = & 
     -\nu \sum_{j} m_{j}K_{\mathrm{mix}, aj}\left(\frac{\mathbf{v}_{aj}\cdot\mathbf{r}_{aj}}{\left|{\mathbf{r}_{aj}}\right|^{2} + \eta^{2}}\right)\mathbf{r}_{aj}W_{aj}(h_{a}), \\
      \left(\frac{\mathrm{d}\mathbf{v}_{\beta}}{\mathrm{d}t}\right)_{\mathrm{coll}} = & +\nu \sum_{b} m_{b}K_{\mathrm{mix},bi}\left(\frac{\mathbf{v}_{bi}\cdot\mathbf{r}_{bi}}{\left|{\mathbf{r}_{bi}}\right|^{2} + \eta^{2}}\right)\mathbf{r}_{bi}W_{ib}(h_{b}).
\end{split}
\end{align}
In this context, $\eta = 0.001\;h^{2}$ \citep{MONAGHAN1995} is a small regularization parameter introduced to avoid singularities in the denominator, while $\nu$ denotes the reciprocal of the number of spatial dimensions.\\
The challenge with the formal SPH approach presented in Eq. \eqref{collisional sph} lies in the fundamentally different nature between $\mathrm{K_{\mathrm{drag}}}$ and the mixing coefficient $\mathrm{K_{\mathrm{mix}}}$. The latter is typically much larger because of the presence of small molecular masses in the denominator. Consequently, the collisional timestep -- which scales inversely with the mixing term, as discussed in Section~\ref{collisional_timestep} -- can be 10 to 100 times smaller than the typical hydrodynamical timestep, or even more in some cases. This makes integration with an explicit method impractical. 
Implicit and semi-implicit schemes for stiff interaction terms have been widely developed in the SPH literature, particularly for gas–dust coupling \citep{Monaghan1997, LaibePrice2014a, Aguilar2014, STOYANOVSKAYA2018, MONAGHAN2020}. These methods are designed to overcome the restrictive timestep constraints introduced by strong drag interactions. While implicit integration would be a formally robust approach to address the stiffness arising from the collisional term, it requires computing the Jacobian of the system's right-hand side, which can be expensive when dealing with a large number of SPH particles.
In our case, to balance accuracy, stability, and computational feasibility, we adopt a splitting approach that separates the hydrodynamical evolution from the collisional relaxation step. This technique offers an efficient alternative that avoids severe timestep constraints.\\
Initially, we solve the hydrodynamical part using the standard SPH formalism, obtaining two velocities, $\mathbf{v_{\alpha}}^{*}$ and $\mathbf{v_{\beta}}^{*}$, and two energies, $\epsilon_{\mathrm{tk},\alpha}^{*}$ and $\epsilon_{\mathrm{tk},\beta}^{*}$, at the end of the timestep. Subsequently, we solve the exponentially relaxing collisional system and correct the original pure hydrodynamical velocities and energies with the new collisional terms. This approach corresponds to the well-known Lie-Trotter splitting method \citep{Trotter1959}.\\
If we neglect the pressure terms and define the velocity difference between the two gases as $\Delta\mathbf{v}= \mathbf{v}_{\alpha} - \mathbf{v}_{\beta}$, the system in Eq. \eqref{pressure_coll_system_simple_form} simplifies to the following differential equation for the velocity difference:
\begin{equation}
    \frac{\mathrm{d}\Delta \mathbf{v}}{\mathrm{d}t} 
     = - (\rho_{\alpha} + \rho_{\beta}) K_{\mathrm{mix}}\Delta \mathbf{v}.
\end{equation}
Assuming that the term $(\rho_{\alpha} + \rho_{\beta}) K_{\mathrm{mix}}$ remains constant over the timestep -- which is a reasonable approximation if $\mathrm{\Delta}t$ is appropriately chosen, for instance, using a CFL condition -- the equation admits an exact solution. Applied over a timestep $\mathrm{\Delta}t$, this solution takes the form:
\begin{equation}
\label{deltaV}
    \Delta\mathbf{v}^{n+1} =  \Delta\mathbf{v^{*}}^{n+1} e^{- (\rho_{\alpha} + \rho_{\beta}) K_{\mathrm{mix}}\mathrm{\Delta}t},
\end{equation}
where $\Delta\mathbf{v^{*}} = \mathbf{v}_{\alpha}^{*} - \mathbf{v}_{\beta}^{*}$. Now, defining the center of mass velocity as
\begin{equation}
\label{vCM}
    \mathbf{v}_{\mathrm{CM}} = \frac{\rho_{\alpha}\mathbf{v}_{\alpha}^{*} + \rho_{\beta}\mathbf{v}_{\beta}^{*}}{\rho_{\alpha} + \rho_{\beta}},
\end{equation}
it is possible to show that the individual gas velocities at the timestep $n+1$ are given by
\begin{align}
\label{velocity_update}
\begin{split}
     \mathbf{v}_{\alpha}^{n+1} = & \;\mathbf{v}_{\mathrm{CM}} + \frac{\rho_{\beta}}{\rho_{\alpha} + \rho_{\beta}}\Delta\mathbf{v}^{n+1}, \\
    \mathbf{v}_{\beta}^{n+1} = & \;\mathbf{v}_{\mathrm{CM}} - \frac{\rho_{\alpha}}{\rho_{\alpha} + \rho_{\beta}}\Delta\mathbf{v}^{n+1}.
\end{split}
\end{align}
At this point, substituting the relations \eqref{deltaV} and \eqref{vCM} in the equations \eqref{velocity_update}, we get
\begin{align}
\label{velocity_update_final_expr}
\begin{split}
     \mathbf{v}_{\alpha}^{n+1} = &  X_{\alpha} \mathbf{v}_{\alpha}^{*} + X_{\beta} \mathbf{v}_{\alpha}^{*} e^{- (\rho_{\alpha} + \rho_{\beta}) K_{\mathrm{mix}}\mathrm{\Delta}t} + X_{\beta} \mathbf{v}_{\beta}^{*}\left(1-e^{- (\rho_{\alpha} + \rho_{\beta}) K_{\mathrm{mix}}\mathrm{\Delta}t}\right), \\
     \mathbf{v}_{\beta}^{n+1} = &  X_{\beta} \mathbf{v}_{\beta}^{*} + X_{\alpha} \mathbf{v}_{\beta}^{*} e^{- (\rho_{\alpha} + \rho_{\beta}) K_{\mathrm{mix}}\mathrm{\Delta}t} + X_{\alpha} \mathbf{v}_{\alpha}^{*}\left(1-e^{- (\rho_{\alpha} + \rho_{\beta}) K_{\mathrm{mix}}\mathrm{\Delta}t}\right),
\end{split}
\end{align}
where $X_{\alpha} = \rho_{\alpha} / (\rho_{\alpha} + \rho_{\beta})$, $X_{\beta} = \rho_{\beta} / (\rho_{\alpha} + \rho_{\beta})$, and $\rho_{\mathrm{tot}} = \rho_{\alpha} + \rho_{\beta}$.\\
A similar approach can be applied to the energy equations. Under the assumption that $f_{\alpha\beta}$ and $\hat{\epsilon}$ remain constant over the timestep, the energy of each gas can be updated by incorporating the corresponding collisional term as follows:
\begin{align}
\label{sph_energy_correction}
\begin{split}
    \epsilon_{\mathrm{tk},\alpha}^{n+1} & = \hat{\epsilon}_{\alpha} + \left(\epsilon_{\mathrm{tk},\alpha}^{*} - \hat{\epsilon}_{\alpha}\right)e^{-f_{\alpha\beta}\; \mathrm{d}t},\\[10pt]
     \epsilon_{\mathrm{tk},\beta}^{n+1}  & = \hat{\epsilon}_{\beta} + \left(\epsilon_{\mathrm{tk},\beta}^{*} - \hat{\epsilon}_{\beta}\right)e^{-f_{\beta\alpha}\; \mathrm{d}t}.
\end{split}
\end{align}
From these expressions, it can be observed that if diffusion is strong ($f \to \infty$), the energies tend toward $\hat{\epsilon}$. Indeed, when the two gases are fully mixed, the temperature and velocity differences vanish: $\Delta T \to 0$, $\Delta \mathbf{v} \to 0$, the velocities approach the center-of-mass velocity, the temperature converges to that of the mixture, and $\hat{\epsilon}$ represents the equilibrium energy value for each gas. Conversely, in the limit $f \to 0$, the energies remain equal to the hydrodynamical values, as no mixing occurs.\\
To incorporate equations \eqref{velocity_update_final_expr} and \eqref{sph_energy_correction} into our SPH scheme, it is essential to determine each term appropriately. The density of each gas is computed using the standard SPH interpolation \citep{Monaghan2005, LP2012b}:
\begin{align}
     \rho_{a}= \sum_{b} m_{b}W_{ab}(h_{a}) \qquad
     \rho_{i}= \sum_{j} m_{j}W_{ij}(h_{i}).
\label{eq:SummDensity_SPH}
\end{align}
Similarly, the local density of gas $\beta$ around an SPH particle of gas $\alpha$, and vice versa, the local density of gas $\alpha$ around an SPH particle of gas $\beta$, are computed as follows: 
\begin{equation}
    \rho_{\mathrm{loc},a}=\sum_{j}^{N_{neigh,\beta}} m_{j}W_{aj}(h_{a}) \qquad
    \rho_{\mathrm{loc},i}=\sum_{b}^{N_{neigh,\alpha}} m_{b}W_{ib}(h_{i}).
    \label{eq:loc_density}
\end{equation}
Therefore,  each SPH particle perceives a total density ($\rho_{\mathrm{tot}}$) that consists of the sum of the density of particles of the same type and the local density contribution from the other component surrounding that particle. Given these quantities, the values of $X_{\alpha}$ and $X_{\beta}$ follow directly from their definitions. The last critical terms in Eq. \eqref{velocity_update_final_expr} are the velocities of the other component, which appear in the third term of each summation. These velocities are estimated by computing the averaged mean velocity of the surrounding particles of the other gas component for each SPH particle:
\begin{align}
    \mathbf{v^{*}}_{\mathrm{loc,a}} =  \frac{\sum_{j} \mathbf{v}_{j}^{*}W_{aj}}{\sum_{j} W_{aj}} \qquad
    \mathbf{v^{*}}_{\mathrm{loc,i}} =  \frac{\sum_{b} \mathbf{v}_{b}^{*}W_{ib}}{\sum_{b} W_{ib}}.
\end{align}
Here, $\mathbf{v^{*}}_{\mathrm{loc,a}}$ and $\mathbf{v^{*}}_{\mathrm{loc,i}}$ are respectively the SPH forms of $\mathbf{v}_{\beta}^{*}$ and $\mathbf{v}_{\alpha}^{*}$ that appear in the last summation terms in Eq. \eqref{velocity_update_final_expr}. Finally, the computation of the collisional energies $\hat{\epsilon}$ in Eq. \eqref{sph_energy_correction} also relies on SPH interpolation, since these quantities implicitly depend on $\Delta T$ and $\Delta\mathbf{v}$, and thus on the local neighborhood of particles from the other gas surrounding each SPH particle.
In particular, we found that employing number-weighted SPH interpolation specifically for the exchange terms improves the accuracy of thermal relaxation and energy conservation in high mass ratio mixtures, such as Neon–Xenon systems.

\section{Numerical Tests}
The approach proposed in Sect.\ref{subsec:atmospheric effects} requires validation, which is addressed through a series of numerical tests presented in this section. First, we will demonstrate the effectiveness of Trotter splitting by comparing it with the standard SPH implementation. Subsequently, we will conduct tests in a controlled environment to verify that the simulations yield results fully consistent with the expected hydrodynamic and thermodynamic behavior. 
To ensure a controlled environment with well-defined equilibration conditions, all tests are performed within a small, closed cylindrical domain with radius and height both set to $1\;\mathrm{cm}$. 
This specific geometry is motivated by two main considerations. First, the small domain size and the choice of equal height and radius promote faster mixing and help minimize boundary effects. Second, as discussed in the final part of this work, the proposed approach is designed to be applicable to the ExoMars ESA programme that will explore the surface of Mars. The rover of this mission is equipped with a drilling system whose drill rod has a diameter of approximately $25\;\mathrm{mm}$ \citep{vago}, closely matching the scale of the test domain used in this study.\\
All SPH simulations are carried out using the open-source Python framework PySPH \citep{Ramachandran2021}.

\subsection{Validation of Trotter splitting: comparison with full SPH evolution}
\label{SPH_splitting_compare}

\begin{figure}
    \centering
    \includegraphics[width=\linewidth]{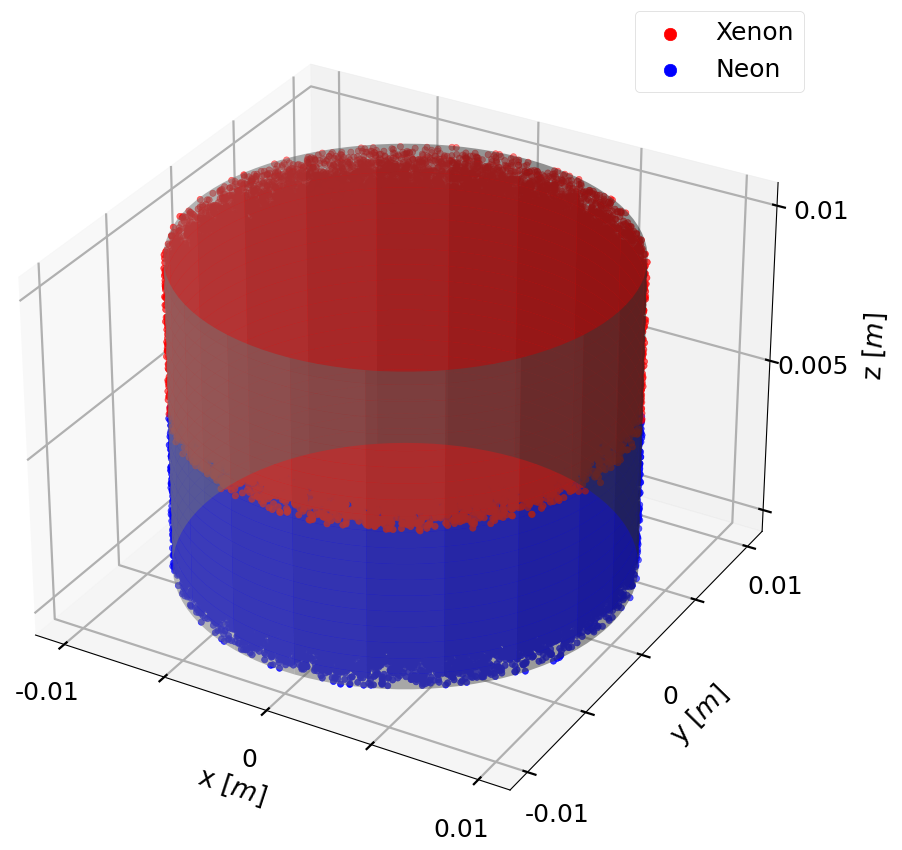}
    \caption{Illustration of the initial system setup. Xenon occupies the upper half of the cylinder, while Neon occupies the lower half.}
    \label{fig:initial_cond}
\end{figure}

\begin{figure}
    \centering
    \includegraphics[width=0.48\textwidth]{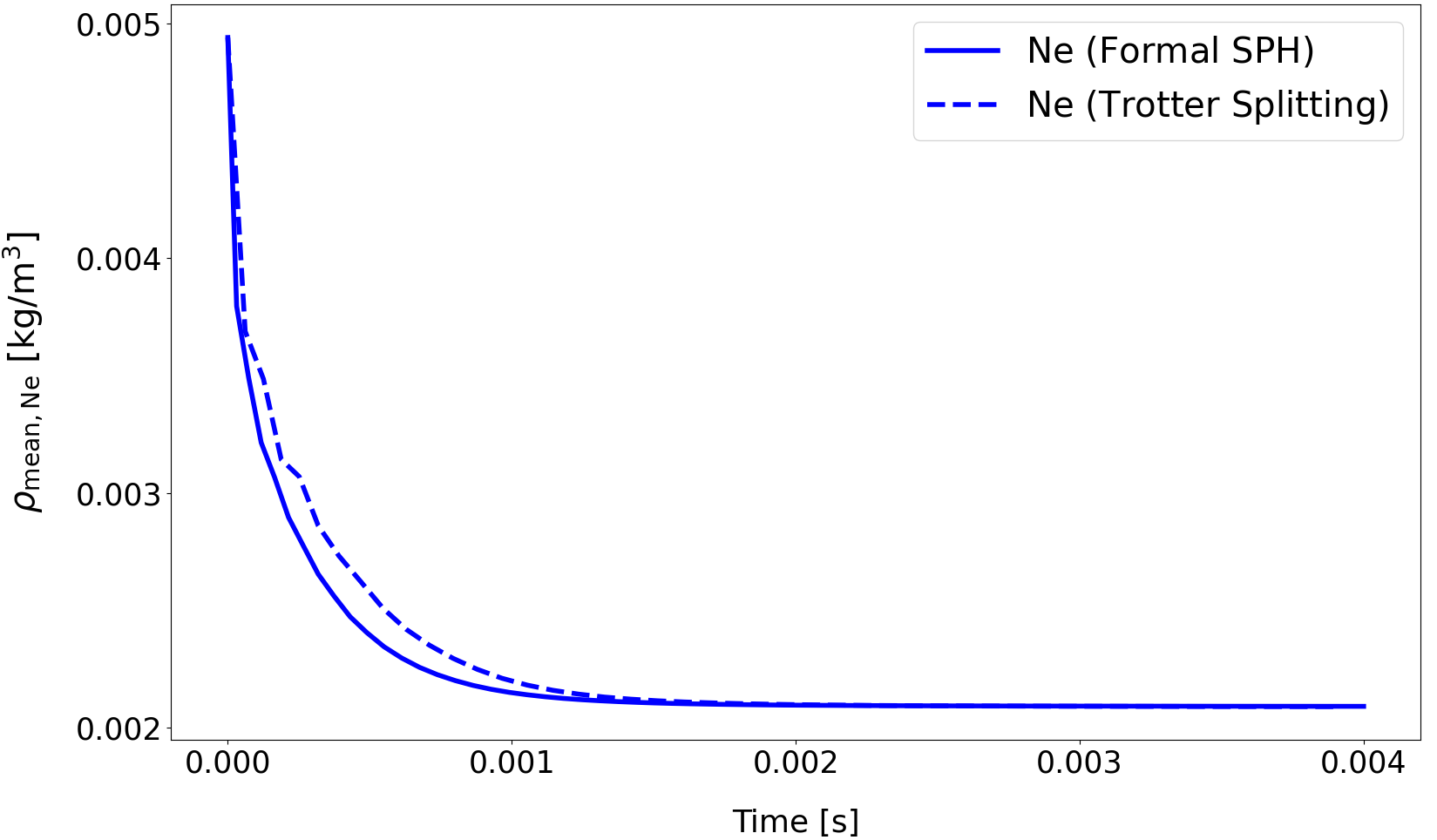}
    \hfill
    \includegraphics[width=0.48\textwidth]{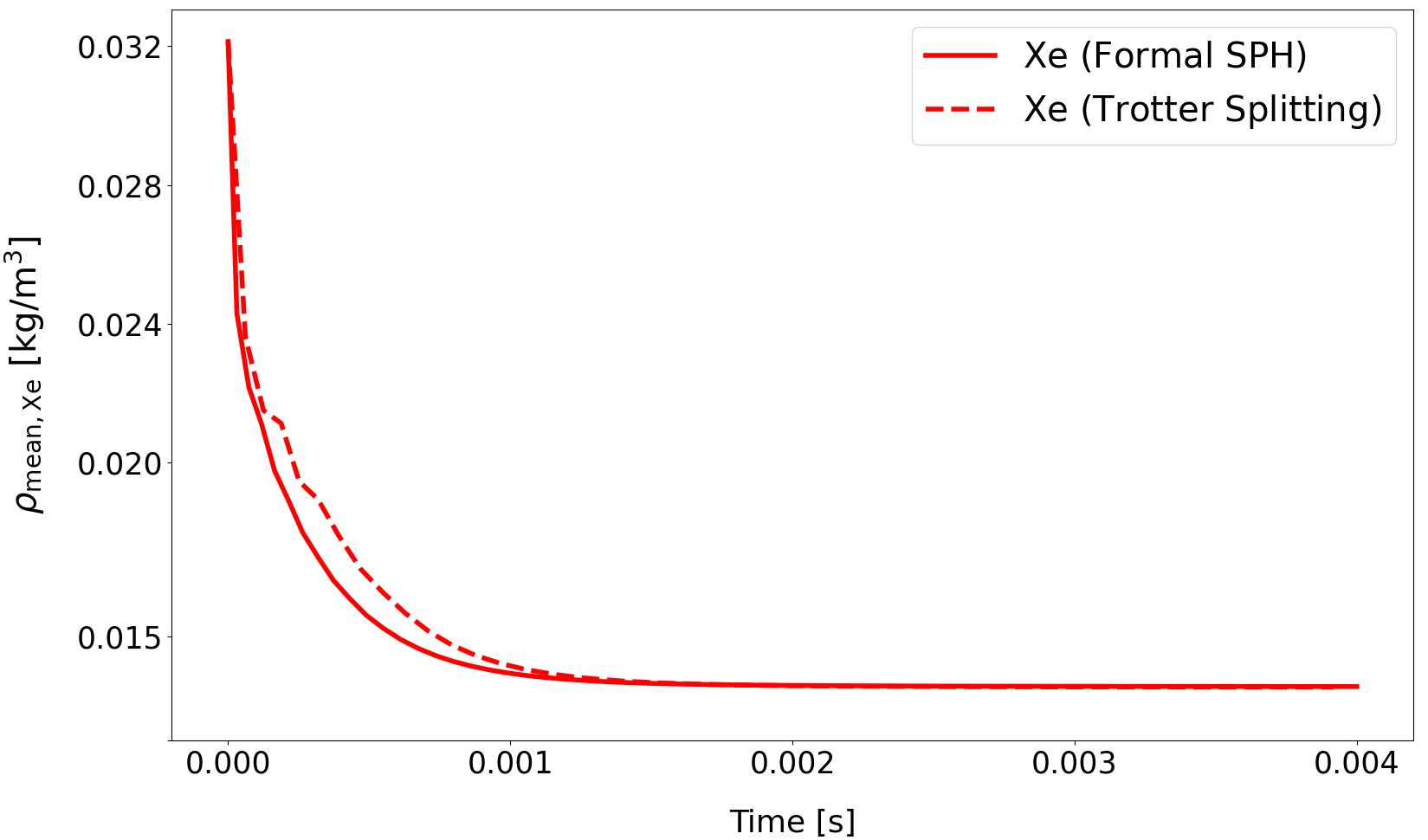}
    \caption{Time evolution of the mean densities for neon (top) and xenon (bottom). Solid lines represent the trend obtained using the formal SPH collisional term from Eq. \eqref{collisional sph}, while dotted lines correspond to the Trotter splitting approach. The discrepancy between the two methods is at most $6\%$ and they show the same asymptotic behavior.}
    \label{fig:sphVStrotter}
\end{figure}

An important first test involves comparing the results obtained using the formal SPH formulation, as detailed in Eq. \eqref{collisional sph}, with those derived from the Trotter splitting method.\\
For this purpose, we consider two monatomic gases with significantly different molecular weights: neon ($m_{Ne} = 20.18\;\mathrm{u}$) and xenon ($m_{Xe} = 131.29\;\mathrm{u}$). Both gases are initially set at a pressure of $P = 610\;\mathrm{Pa}$ and a fixed temperature of $T = 300\;\mathrm{K}$. They are treated as ideal gases with an adiabatic index of $\gamma = 1.66$. Xenon is initially placed in the upper half of the cylindrical domain, while neon occupies the lower half. Fig.\ref{fig:initial_cond} illustrates this initial configuration, providing a visual reference for the simulation setup. Each SPH particle is initialized with the temperature, pressure, and density of the corresponding gas.
\\
This setup is designed specifically to assess the validity and performance of the splitting approach. Accordingly, temperature is held constant throughout the simulation, and only the equations governing density and momentum are evolved; the energy equation is not included in this analysis. Additionally, external forces such as gravity are neglected.
As illustrated in Fig.\ref{fig:sphVStrotter}, which shows the time evolution of mean densities for both gas species, the two approaches yield highly consistent results, with a maximum discrepancy of only $6\%$. Both simulations use $5 \times 10^4$ particles and are fully parallelized using OpenMP on an 18-core workstation. The formal SPH simulation required approximately 6 hours to complete, while the Trotter splitting method finished in about 40 minutes. This substantial reduction in computational time highlights the efficiency of the Trotter approach without compromising the accuracy of the results.\\
Finally, to further support the accuracy of the numerical implementation, the DUSTYBOX test \citep{laibe11}, performed in both non-stiff and stiff regimes, is presented in Appendix \ref{sec:dustybox}.

\subsection{Hydro-thermodynamic validation in a controlled mixture}
\label{Argon_Krypton}

In this second test, we consider two monoatomic gases, argon ($m_{Ar} = 39.948\;\mathrm{u}$) and krypton ($m_{Kr} = 83.798\;\mathrm{u}$), selected because their molecular weight ratio closely resembles that of water vapor and carbon dioxide, which are the gases of interest for future applications in realistic Mars scenarios.\\
The initial configuration mirrors that of the previous test, with krypton occupying the upper half of the closed cylindrical domain and argon positioned in the lower half. Both gases are initialized at the same pressure ($610 \; \mathrm{Pa}$), but with different temperatures: $400 \; \mathrm{K}$ for argon and $300 \; \mathrm{K}$ for krypton. As in the previous case, no external forces are considered. However, in this simulation, we solve the full set of hydrodynamic equations, including the energy equation, allowing for energy exchange and thermal equilibration between the two gases. The results presented in this section are obtained from a SPH simulation using $2 \times 10^{5}$ particles. Here, we employ a larger number of particles than in the previous test. This is because in the earlier comparison, the formal SPH scheme, which is limited by a much smaller collisional timestep, was computationally expensive even with relatively few particles.\\
At this stage, since the two gases are confined within a closed cylindrical domain, the final equilibrium temperature can be predicted using the weighted average:
\begin{equation}
\label{equilibrium_T}
    T_{\mathrm{eq}} = \frac{M_{1}\;c_{\mathrm{v},1}\; T_{1} + M_{2}\;c_{\mathrm{v},2}\; T_{2}}{M_{1}\;c_{\mathrm{v},1} + M_{2}\;c_{\mathrm{v},2}},
\end{equation}
where $M_{i}$, $c_{\mathrm{v}, i}$, and $T_{i}$ represent the mass, specific heat at constant volume, and initial temperature of each gas, respectively. For Ar-Kr system, we have that $c_{\mathrm{v}, Ar} = 312 \; \mathrm{J/(kg \; K)}$ and  $c_{\mathrm{v}, Kr} = 151 \; \mathrm{J/(kg \; K)}$. These values are obtained exploiting the standard hydrodynamic relation $c_{\mathrm{v}, i} = (3 \; k_{B}) / (2 \; m_{i})$. Moreover, because the total mass is conserved and the volume occupied by each gas doubles during the mixing process, the final equilibrium densities are expected to be half of their initial values.
\begin{figure}
    \centering
    \includegraphics[width=\linewidth]{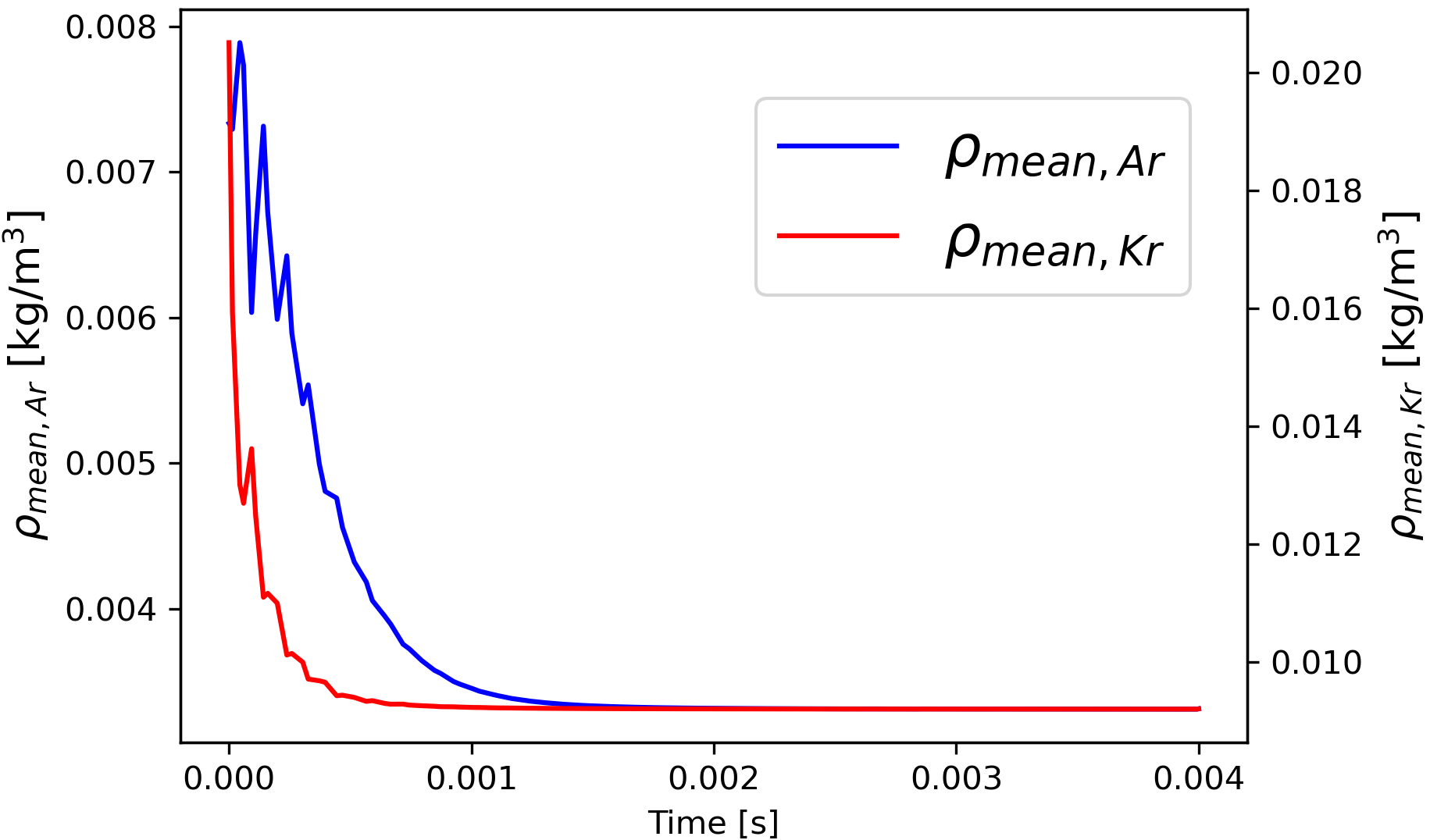}
    \caption{Time evolution of the mean densities of argon (blue) and krypton (red) during the mixing process.}
    \label{fig:mean_densities_Ar_Kr}
\end{figure}\\
As shown in Fig.\ref{fig:mean_densities_Ar_Kr}, both gases indeed approach an equilibrium density slightly below half their starting values.  The simulation thus demonstrates the correct behavior, although the final mean density is underestimated by approximately $8-9\%$ relative to the theoretical prediction. This underestimation error decreases as the number of SPH particles increases. As shown in Fig.\ref{fig:sphVStrotter}, with $5 \times 10^{4}$ the discrepancy exceeds $25\%$.
Increasing the resolution leads to a progressive reduction in the error, which stabilizes around $7-9\%$, even when using up to $3 \times 10^{5}$ particles. This persistent small discrepancy is primarily attributed to the influence of boundary conditions in the SPH framework rather than to the implementation of the collisional terms. In these tests, we employ a simple mirroring scheme in cylindrical coordinates to reflect particles at the domain boundaries \citep{Colagrossi+2009}. While this method ensures confinement, it introduces inaccuracies due to kernel truncation effects near the boundaries, which typically result in an underestimation of the density close to the cylinder walls. 
All simulations presented in this work use the standard cubic spline kernel \citep{Monaghan2005}. To assess the sensitivity of the results to the kernel choice, we also performed an additional test using a Wendland $C^{6}$ kernel \citep{Wendland1995}. This test produced consistent density evolution, indicating that the results are not significantly affected by the choice of kernel function, as expected (see Appendix \ref{sec:kernel}). Further confirmation that density bias arises from boundary effects -- and not from the collisional model itself -- comes from Appendix \ref{sec:Ideal_gas_mix}, where we evolve two gases without collisions, solving only the well known Euler equations in SPH form \citep{Monaghan2005}. Despite the absence of interaction terms, a similar underestimation of the mean densities is observed, reinforcing the conclusion that the error is intrinsic to the SPH boundary treatment.\\
Moreover, always looking at Fig.\ref{fig:mean_densities_Ar_Kr}, we can observe that the heavier gas (krypton) reaches its equilibrium density faster than the lighter argon. This behavior is driven by the pressure gradient at the interface, which causes krypton to displace into the argon region more rapidly. In contrast, the lighter argon initially experiences mechanical compression, leading to a transient increase in its density before gradually decompressing and redistributing. This dynamic delays its full equilibration. 
\begin{figure}
    \centering
    \includegraphics[width=\linewidth]{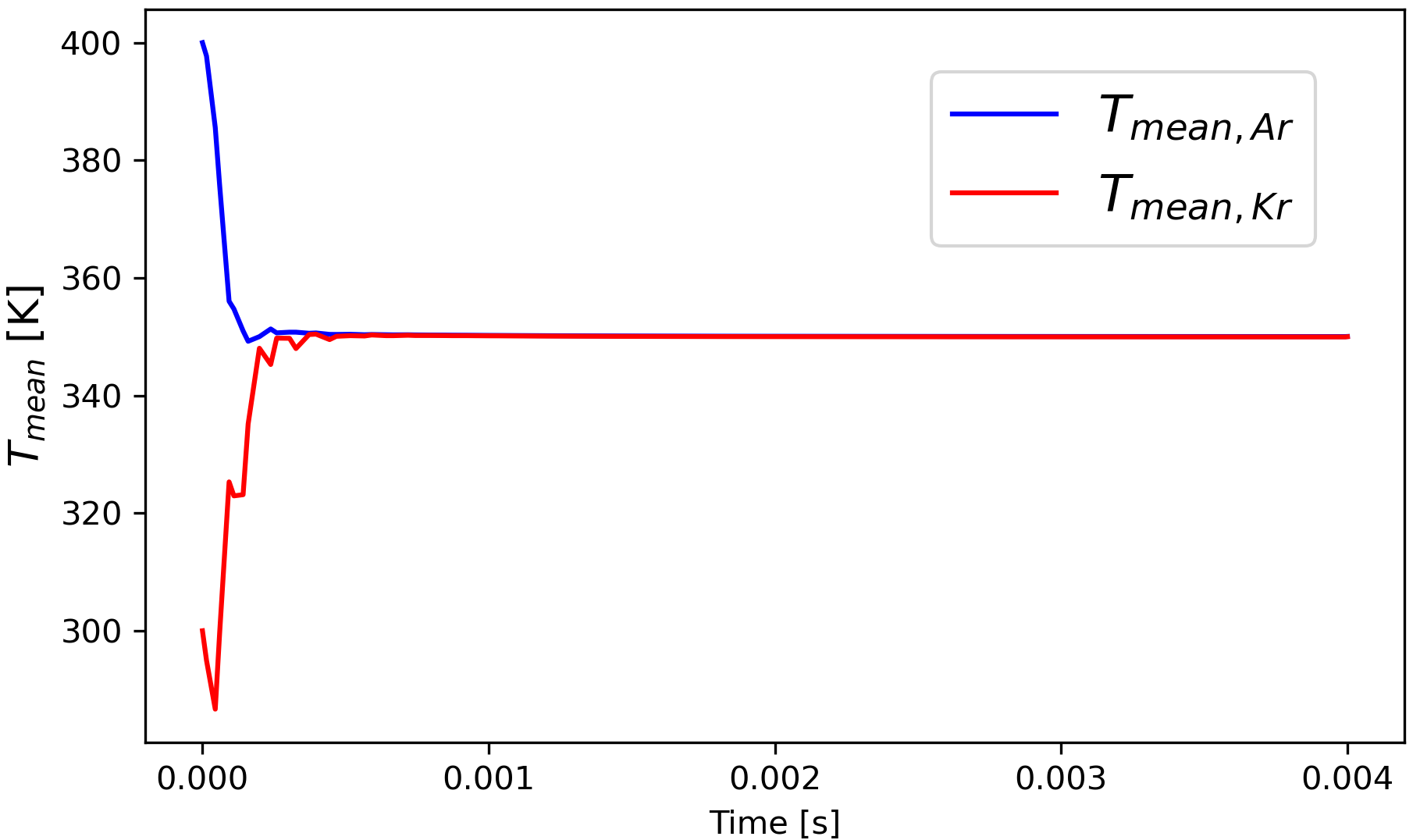}
    \caption{Time evolution of the mean temperature of argon (blue) and krypton (red) during the mixing process.}
    \label{fig:mean_T_Ar_Kr}
\end{figure}
Regarding the evolution of temperature, Fig.\ref{fig:mean_T_Ar_Kr} shows that the system relaxes to a common equilibrium temperature of approximately $349 \mathrm{K}$, in good agreement with the theoretical value of $T_{\mathrm{eq}} \approx 343 \mathrm{K}$, yielding a relative error below $2\%$. For comparison, in Appendix \ref{sec:Ideal_gas_mix} we report the ideal gas limit (no collisions), where temperature equilibration is absent, as expected. Notably, temperature relaxation precedes density equilibration, as expected from physical considerations: energy is exchanged directly through collisions, while mass redistribution is slower.
\begin{figure}
    \centering
    \includegraphics[width=\linewidth]{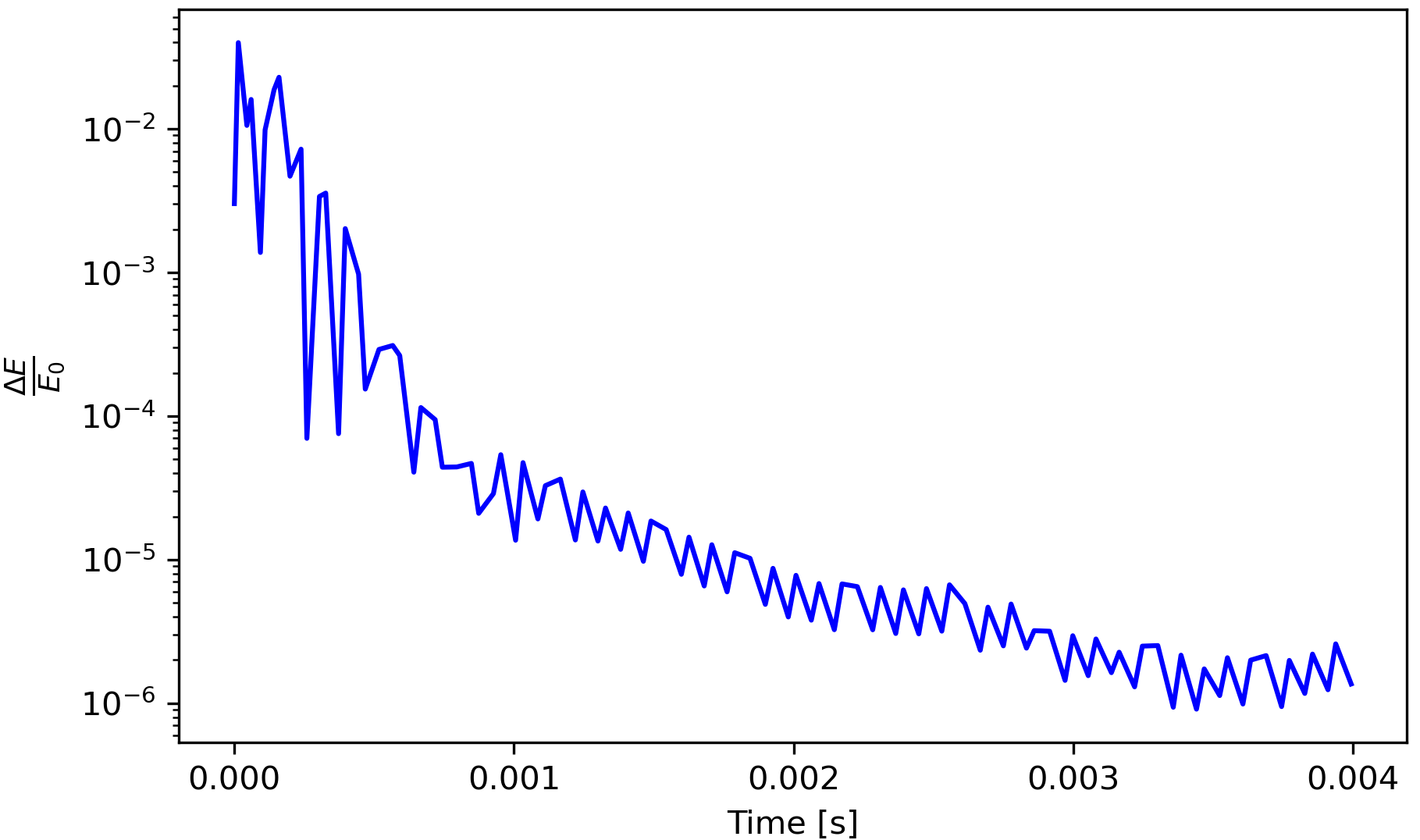}
    \caption{Change in total energy over time, normalized by the initial energy.}
    \label{fig:energy_conservation}
\end{figure}
Finally, the simulation conserves both total momentum and energy with good accuracy. As reported in Fig.\ref{fig:energy_conservation} the relative error in total energy remains of the order of $10^{-5}$ throughout the simulation, except for a brief initial transient phase. In addition, the model satisfies the second law of thermodynamics. Indeed, as demonstrated in Appendix \ref{sec:entropy}, the entropy of the system increases over time, consistently with the irreversible nature of the mixing process.\\
However, accurately reproducing the equilibrium values alone is not sufficient to validate the model. As demonstrated in Appendix \ref{sec:Ideal_gas_mix}, even in the absence of collisions, the mean densities of the two gases tend to converge toward the same equilibrium value. This highlights the importance of not only obtaining the correct final state, but also ensuring that the system relaxes on a physically meaningful timescale. When collisional effects are neglected, the gases appear to mix almost instantaneously, which contradicts the expected physical behavior.
To give an estimate of the expected mixing time, we first need to determine the diffusion coefficient of the binary gas mixture. According to \cite{Hudson2007}, for a Lennard-Jones potential, the diffusion coefficient $D_{\alpha\beta}$ is given by
\begin{equation}
    D_{\alpha\beta} = \frac{3}{8\sigma_{\alpha\beta}^{2}}\frac{k_{B}T_{\mathrm{mix}}}{P_{0}}\sqrt{\frac{k_{B}T_{\mathrm{mix}}}{2\pi \;m_{\alpha\beta}}}\frac{1}{W_{\alpha\beta}^{(1,1)}},
\end{equation}
where $P_{0}$ is the initial pressure. This expression is a first-order approximation derived from Chapman-Enskog theory \citep{Chapman1970}. At the lowest order, the diffusion coefficient is independent of the relative concentration of the two gases, leading to the symmetric relation $D_{\alpha\beta} = D_{\beta\alpha} \equiv \mathrm{D}$. Therefore, using the Lennard-Jones parameters \citep{Oh2013} reported in Table \ref{table2:epsilon&sigma_Ar_Kr}, we find that the Ar-Kr diffusion coefficient is $\mathrm{D} \approx 3.4\times10^{-3}\; \mathrm{m^{2}/s}$.
\begin{table}
\caption{Lennard-Jones parameters for Ar and kr.}
\centering
\begin{tabular}{|c|c|c|}
\hline
Gas & $\epsilon / k_{B}$ [K] & $\sigma$ [$\mathrm{\mbox{\AA}}$] \\
\hline
Ar & 116.81 & 3.401  \\
Kr & 164.56 & 3.601  \\
\hline
\end{tabular}
\label{table2:epsilon&sigma_Ar_Kr}
\end{table}\\
Nevertheless, for estimating the correct equilibrium time we have to consider that the diffusion timescale can be enhanced by making use of a non-flat velocity profile. This is exactly the case of our model, in which velocities are also affected by the density gradient between the two gases. The classical diffusion coefficient is thus replaced by an effective value \citep{Taylor, Aris}:
\begin{equation}
\label{effective_D}
    \mathrm{D_{eff}} = D\left(1 + \frac{1}{48}\mathrm{Pe}^{2}\right).
\end{equation}
The factor $1/48$ is due to the cylindrical geometry, while $\mathrm{Pe}$ is the Péclet number \citep{patankar1980}. It is a dimensionless number which compares the relative importance of advection and diffusion for transport phenomena:
\begin{equation}
    \mathrm{Pe} = \frac{t_{\mathrm{diffusion}}}{t_{\mathrm{advection}}} = \frac{L^{2}/\mathrm{D}}{L/U} = \frac{L\;U}{\mathrm{D}}.
\end{equation}
Here, L is the characteristic length over which diffusion occurs (half of the height of the cylinder domain), while U is the local flow velocity. Although the relation \eqref{effective_D} is formally derived for laminar flow in long cylindrical pipes, we use its effective diffusion correction as a heuristic estimate of flow-enhanced mixing in our closed cylindrical domain. While the geometry differs from the classical case, the relation captures the key interplay between advection and diffusion in confined systems. At this point, to give an estimate of the effective diffusion coefficient, we use for U the peak difference between the mean velocities of the two gas species, occurring at early times during mixing when relative advection is strongest. In this case, $ U \approx 12 \; \mathrm{m/s}$. an equilibrium timescale of about $1.0\times 10^{-3}\; \mathrm{s}$, which, looking at Fig.\ref{fig:mean_densities_Ar_Kr}, is close to the time at which Argon mean density relax to equilibrium.\\
It is important to emphasize that the estimated mixing time discussed here should be regarded as an order of magnitude reference rather than a precise prediction. The key point is that, with the inclusion of collisional terms, the mixing process is no longer instantaneous but occurs over a physically plausible timescale, which is comparable to the expected theoretical estimate. The numerical scheme employed in this work relies on a first-order operator splitting between hydrodynamic evolution and collisional relaxation, combined with an explicit two-steps Euler time integration. While this approach is computationally efficient and yields satisfactory results, future work could explore the use of higher-order time integration methods (e.g., Runge–Kutta schemes) or more accurate operator splitting techniques \citep{Strang} to assess their impact on the overall solution accuracy.

\subsection{Neon-Xenon system}
\label{sec:Neon_xenon}

\begin{figure}
    \centering
    \includegraphics[width=\linewidth]{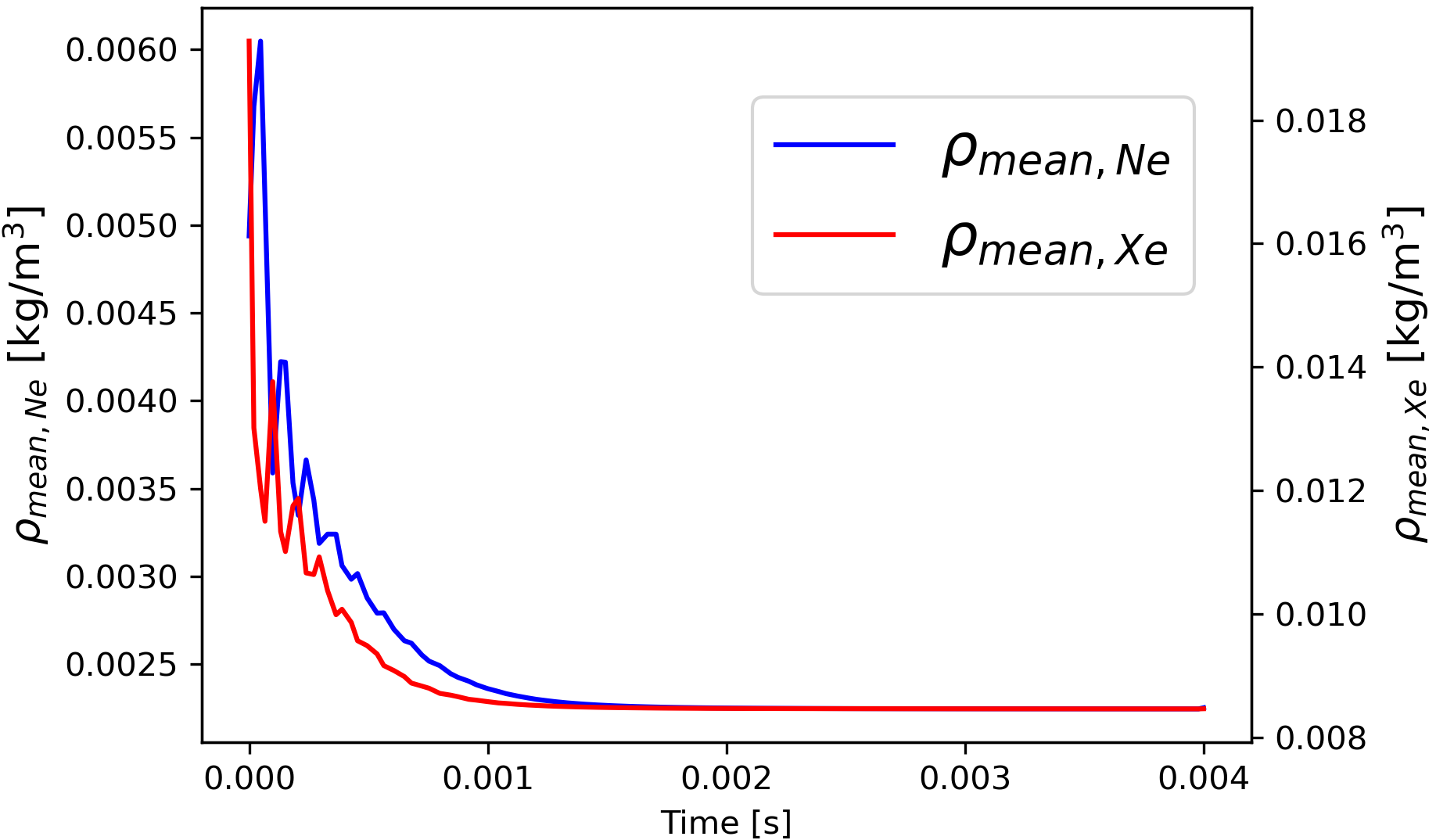}
    \caption{Time evolution of the mean densities of neon (blue) and xenon (red) during the mixing process.}
    \label{fig:densities_ne_xe}
\end{figure}

\begin{figure}
    \centering
    \includegraphics[width=\linewidth]{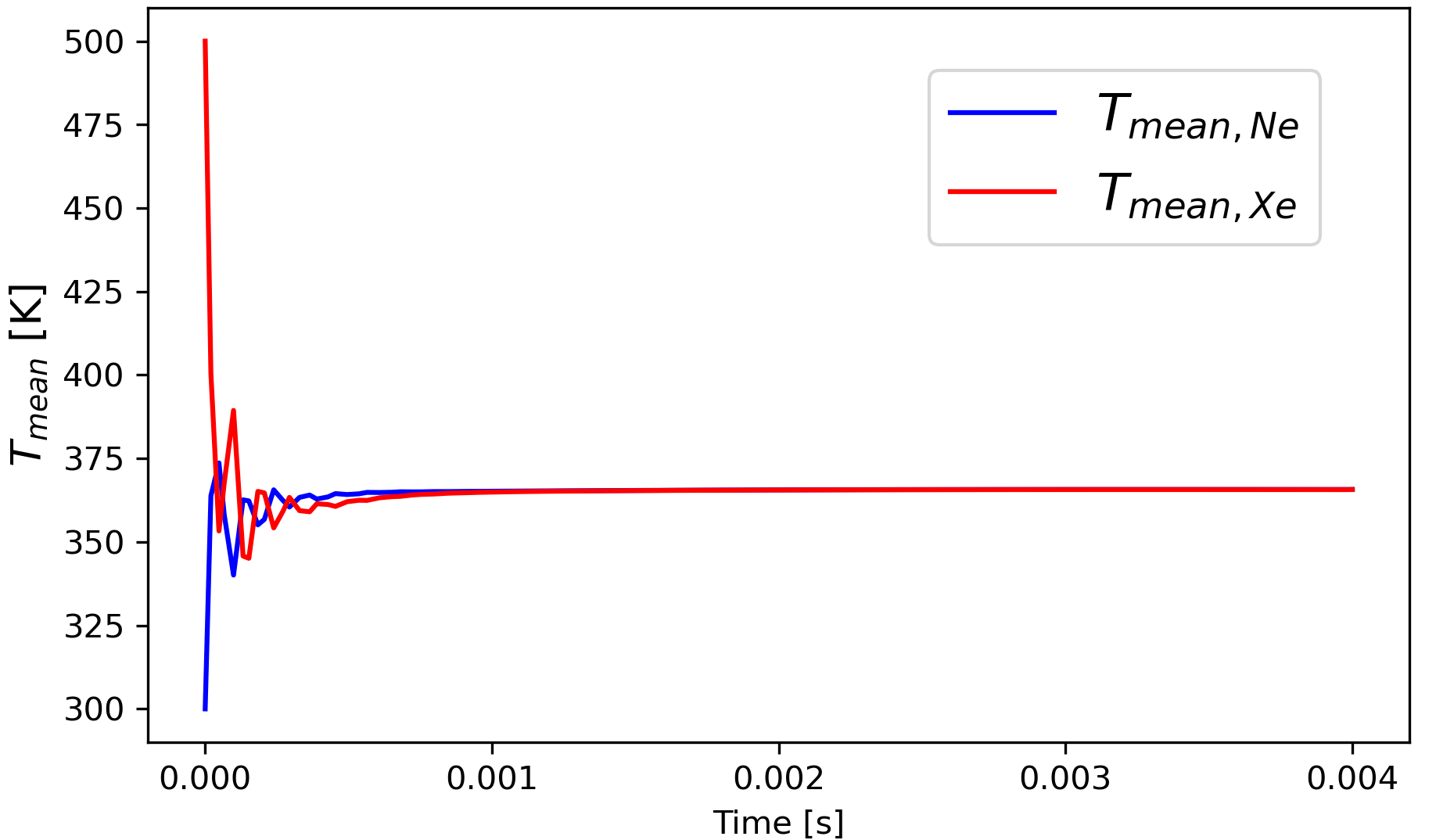}
    \caption{Time evolution of the mean temperatures of neon (blue) and xenon (red) during the mixing process.}
    \label{fig:T_ne_xe}
\end{figure}

As a final test, we examine a gas mixture with a higher molecular mass ratio: xenon and neon. This case is important to confirm that the model performs reliably across a broader range of conditions, not just in the previously discussed scenario. To emphasize this point, we use the same configuration described in Section \ref{SPH_splitting_compare}, but this time xenon is initialized at a temperature of $500 \mathrm{K}$ and neon at $300 \mathrm{K}$. Unlike the earlier case, the heavier gas now starts at the higher temperature, and the initial temperature difference between the two gases is also larger. Table \ref{table2:epsilon&sigma_Ne_Xe} summarizes initial conditions and relevant parameters for the Ne–Xe mixture. Results for this test are again obtained using an SPH simulation with $2\times 10^{5}$ particles.
\begin{table}
\centering
\caption{Lennard-Jones parameters \citep{Oh2013}, specific heat at constant volume, and initial temperature and pressure for Ne and Xe.}
\begin{tabular}{|c|c|c|c|c|c|}
\hline
Gas & $\epsilon / k_{B}$ [K] & $\sigma$ [$\mathrm{\mbox{\AA}}$] & $c_{\mathrm{v}}$ [$\mathrm{J/(Kg \; K)}$] & $T \; [\mathrm{K}]$ & $P\; [\mathrm{Pa}]$\\
\hline
Ne & 36.8 & 2.77  & 618 & 300 & 610\\
Xe & 218.2 & 4.05 & 97 & 500 & 610 \\
\hline
\end{tabular}
\label{table2:epsilon&sigma_Ne_Xe}
\end{table}\\
Looking at Fig.\ref{fig:densities_ne_xe}, we observe that, as in the Ar–Kr case, the lighter gas reaches its equilibrium density slightly more slowly than the heavier one. This behavior is consistent with the expected dynamics, since the lighter gas is initially compressed and takes longer to redistribute. In this Ne–Xe case, the molecular mass ratio is significantly larger and, as a result, the initial mechanical compression of neon is more pronounced than what was seen with argon.  Despite this, both gases ultimately relax toward a uniform spatial distribution, with the final mean densities agreeing with the expected equilibrium values within a relative error of approximately $8-9\%$. Using the Lennard-Jones parameters reported in Table \ref{table2:epsilon&sigma_Ne_Xe}, we estimate the binary diffusion coefficient for the Ne-Xe mixture to be approximately $3.5\times 10^{-3}\; \mathrm{m^{2}/s}$. This is almost the same value as previously calculated for the Argon–Krypton system. Furthermore, the peak difference between the mean velocities of the two gases during the early mixing phase is of the same order of magnitude as in the Ar–Kr case. Since the Péclet number and effective diffusion enhancement scale with this velocity difference, we obtain a comparable mixing timescale for the two systems.
\\
In terms of thermal relaxation, instead, the mean temperatures of both neon and xenon converge toward a common equilibrium value of approximately $366 \; \mathrm{K}$, as shown in Fig.\ref{fig:T_ne_xe}. This value is in very good agreement with the expected theoretical equilibrium temperature of $375 \; \mathrm{K}$ derived from \eqref{equilibrium_T}. The relative error remains around $2\%$, consistent with what was observed in the Ar–Kr case. This confirms that the collisional energy exchange mechanism implemented in the model is capable of capturing thermal equilibration even in the presence of a large molecular mass ratio and temperature asymmetry. Notably, despite the increased complexity of the Ne–Xe interaction, no artificial temperature drift is observed, and the system stabilizes near the correct thermodynamic equilibrium.

\section{Conclusions}

In this work, we present a novel SPH framework for simulating binary gas mixing, with explicit treatment of interspecies collisional momentum and energy exchange based on a kinetic relaxation model. The method is based on a two-step Euler integrator coupled with a first-order operator splitting. While this represents a low-order scheme, our results demonstrate that even this baseline implementation is able to provide physically consistent and quantitatively reasonable results, including realistic thermal equilibration, density relaxation, and energy conservation across a range of test cases. Although all tests were conducted within the same domain geometry, the underlying physics of the model is independent of the boundary conditions and remains generally valid. In particular, the operator splitting approach is applicable in any setting where the diffusion timescale remains substantially shorter than the hydrodynamic timestep. \\
This model becomes particularly important when considering realistic planetary scenarios, 
such as the emission of volatile and refractory material in the Martian environment. Volatile materials generally refer to a mixture of gases (e.g., water vapor, $\mathrm{CO_2}$), while refractory to solid components (e.g., ice, dust). The investigation of these events is particularly intriguing because they may offer clues in the search for biological life and provide insight into a planet’s formation, evolution, and surface–subsurface conditions. Given the significant variability observed in the case of these emissions, where fluids may accelerate to supersonic velocities and densities decrease by several orders of magnitude, the Lagrangian SPH description is one of the most effective modeling methods.\\
The release of volatile material can occur spontaneously or can be triggered by external factors, as in the case of the ESA Rosalind Franklin rover \citep{vago}. Scheduled for launch in 2028 as part of the ExoMars mission, the rover is equipped with a drilling system capable of analyzing down to $2$ meters in the subsurface of Oxia Planum \citep{corradini}. The primary goal of the drill is to collect subsurface samples with significant astrobiological potential for detailed analysis. Before the extraction, the sample will be examined by Ma\_MISS (Mars Multispectral Imager for Subsurface), which will capture hyperspectral images of the borehole's lateral wall created by the drill \citep{mamiss}. 
In future work, we aim to extend our SPH model to investigate how drilling operations might influence the presence and stability of hypothetical volatile materials, such as water ice, in the Martian subsurface. In this context, the gas mixing framework presented here provides a crucial foundation for accurately incorporating atmospheric effects into such models. Indeed, it allows each gas species to be evolved with its own set of Euler equations and interspecies interaction terms. This structure offers a natural and modular pathway for incorporating additional physical processes -- such as drag interactions between gases and ice grains or dust particles \citep{LP2012a} -- in a computationally efficient manner.

\section*{Acknowledgements}

This work was supported by the ASI-INAF grant "Attività scientifica di preparazione all'esplorazione marziana 2023-3-HH.0".

%%%%%%%%%%%%%%%%%%%%%%%%%%%%%%%%%%%%%%%%%%%%%%%%%%
\section*{Data Availability}

No data was used for this article.


\bibliographystyle{mnras}
\bibliography{example}

%%%%%%%%%%%%%%%%% APPENDICES %%%%%%%%%%%%%%%%%%%%%
\appendix

\section{DUSTYBOX tests}
\label{sec:dustybox}
\begin{figure}
    \centering
    \includegraphics[width=\linewidth]{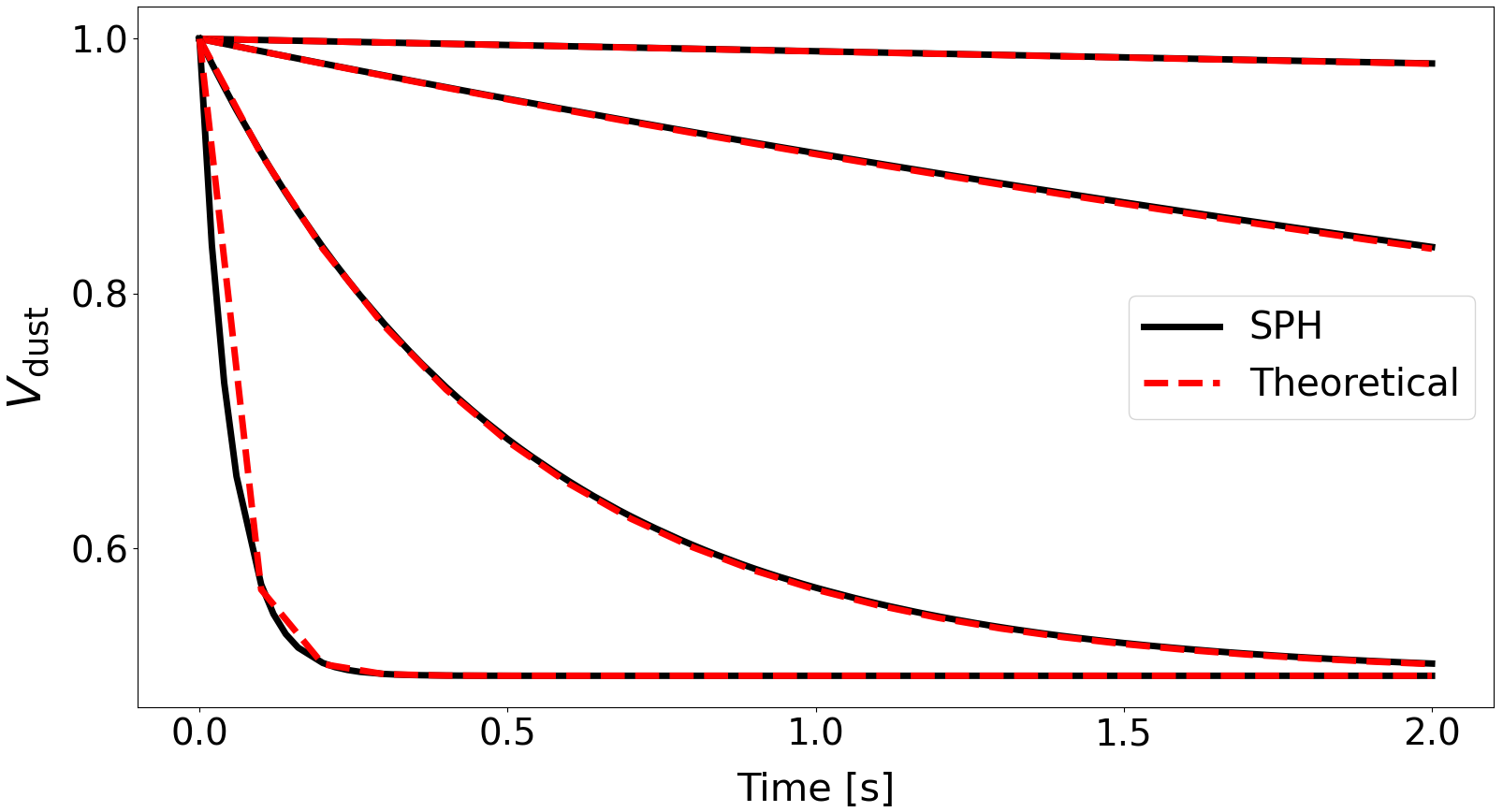}
    \caption{Time evolution of the dust velocity in the DUSTYBOX test with a linear drag term. Results are shown for four different values of the drag coefficient: $K_{0} = 0.01$, $0.1$, $1$, and $10$ (top to bottom lines in the plot).}
    \label{fig:drag_tests}
\end{figure}

\begin{figure}
    \centering
    \includegraphics[width=\linewidth]{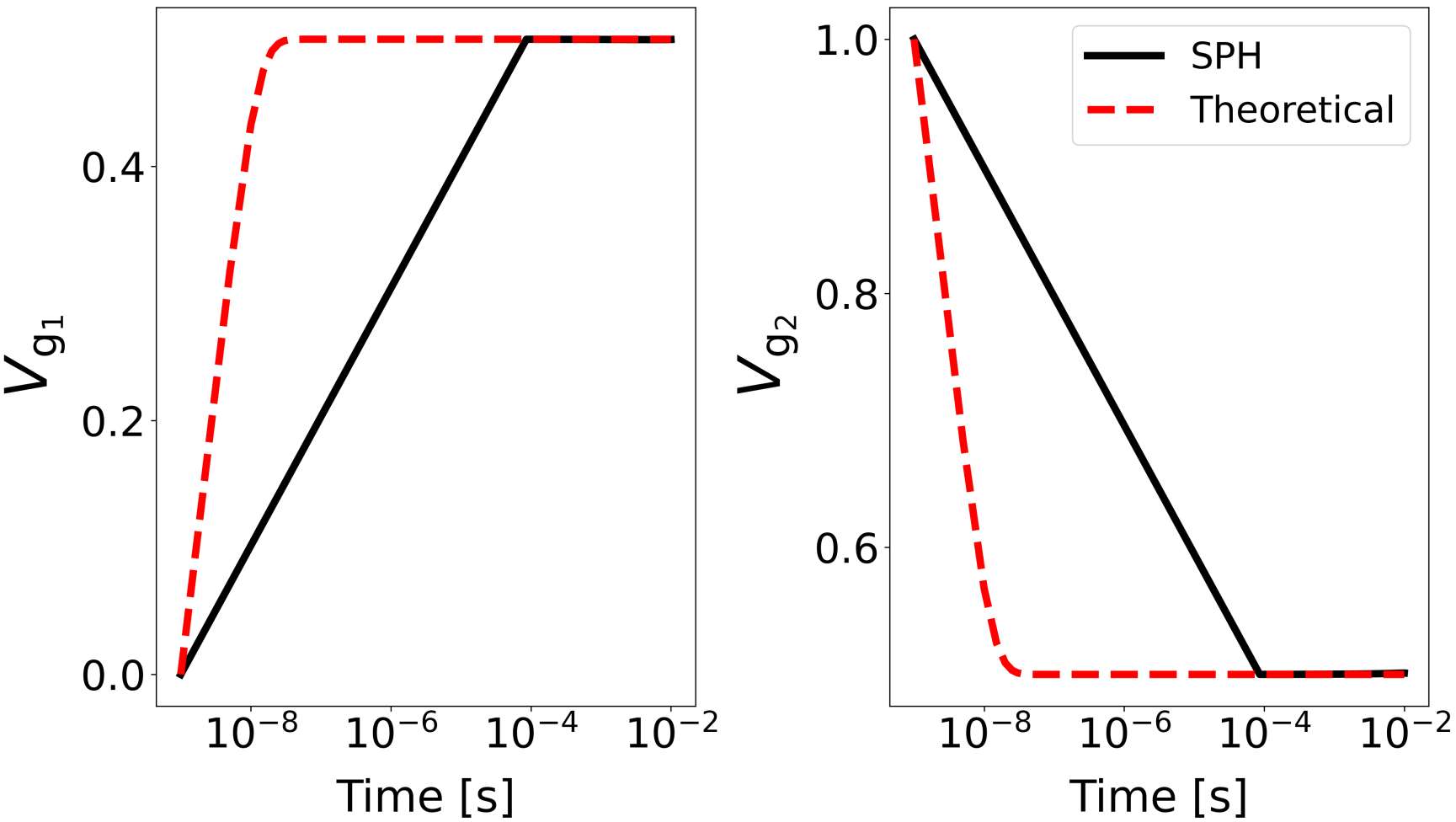}
    \caption{Velocity evolution in the DUSTYBOX test with a strong coupling coefficient $K = 10^{8}$. The left panel shows the velocity of gas 1 ($g_{1}$), initially at rest, and the right panel shows gas 2 ($g_{2}$), initially moving with $v_{x} = 1$.}
    \label{fig:trotter_dusty}
\end{figure}

To validate the implementation of the SPH collisional term and the splitting method introduced in Section~\ref{SPH_splitting_compare}, we performed additional numerical tests using the classical DUSTYBOX setup \citep{laibe11}.\\
Firstly, we verified the correctness of the full SPH implementation presented in Eq.\eqref{collisional sph}. As discussed in the paper, the form of this equation matches that of the drag term presented in \cite{LP2012a, LP2012b}, except for a density term in the denominator. Therefore, to enable a direct comparison with the analytical solutions from \cite{laibe11}, we adapted our expressions (Eqs.\eqref{pressure_coll_system_simple_form} and \eqref{collisional sph}) to the gas–dust context. We adopt the standard DUSTYBOX setup as described in \citet{laibe11}, using a 3D periodic domain $[0,1]^3$ populated with $20^3$ gas particles arranged on a regular lattice and an equal number of dust particles offset by half a lattice spacing. Both gas and dust components have uniform density set to unity; gas particles are initially at rest, while dust particles are assigned a uniform velocity of $v_{d,x} = 1$ along the $x$-direction.
According to the expected regime, the drag coefficient $K_{\rm drag}$ can be either constant or a function of the relative velocity magnitude $|\Delta \mathbf{v}|$, where $\Delta \mathbf{v} = \mathbf{v}_{d} - \mathbf{v}_{g}$ \citep{laibe11}. Nevertheless, since the focus of this work is on gas-gas mixing, rather than solid-gas drag coupling, and because mixing does not exhibit distinct velocity-dependent regimes, we restrict this test to the case of a constant, linear drag coefficient $K_{\rm drag} = K_{0}$. Fig.\ref{fig:drag_tests} shows the evolution of dust velocity compared to analytical solutions, highlighting the good agreement between our results and the theoretical predictions.
\\
We then perform a DUSTYBOX test with a very large friction coefficient, $K = 10^8$, representative of typical values for $K_{\rm mix}$ in binary gas mixing scenarios. This leads to a highly stiff mixing term and provides a suitable framework to apply the splitting scheme presented in the paper. In the new setup, both gas components are initialized with uniform and constant densities and pressures. The initial velocity conditions are the same as before: one gas component, labeled $\mathrm{g_{1}}$, is at rest, while the other gas component ($\mathrm{g_{2}}$) moves with a velocity $v_{x} = 1$. In Fig. \ref{fig:trotter_dusty}, we compare the numerical solution to the analytical one for both fluid components. In this stiff regime, the characteristic decay timescale becomes significantly shorter than the simulation timestep. As a result, the exponential term in the analytical solution causes the velocities to relax to equilibrium almost instantaneously. This relaxation occurs within a single integration step and is not temporally resolved. Nevertheless, the simulation captures the correct asymptotic behavior, with the velocities of the two gas components rapidly converging to their common terminal value. Although the intermediate phase of the velocity decay is not visible due to timestep resolution limits, the final equilibrium state is accurately reproduced.

\section{Kernel function comparison}
\label{sec:kernel}
\begin{figure}
    \centering
    \includegraphics[width=\linewidth]{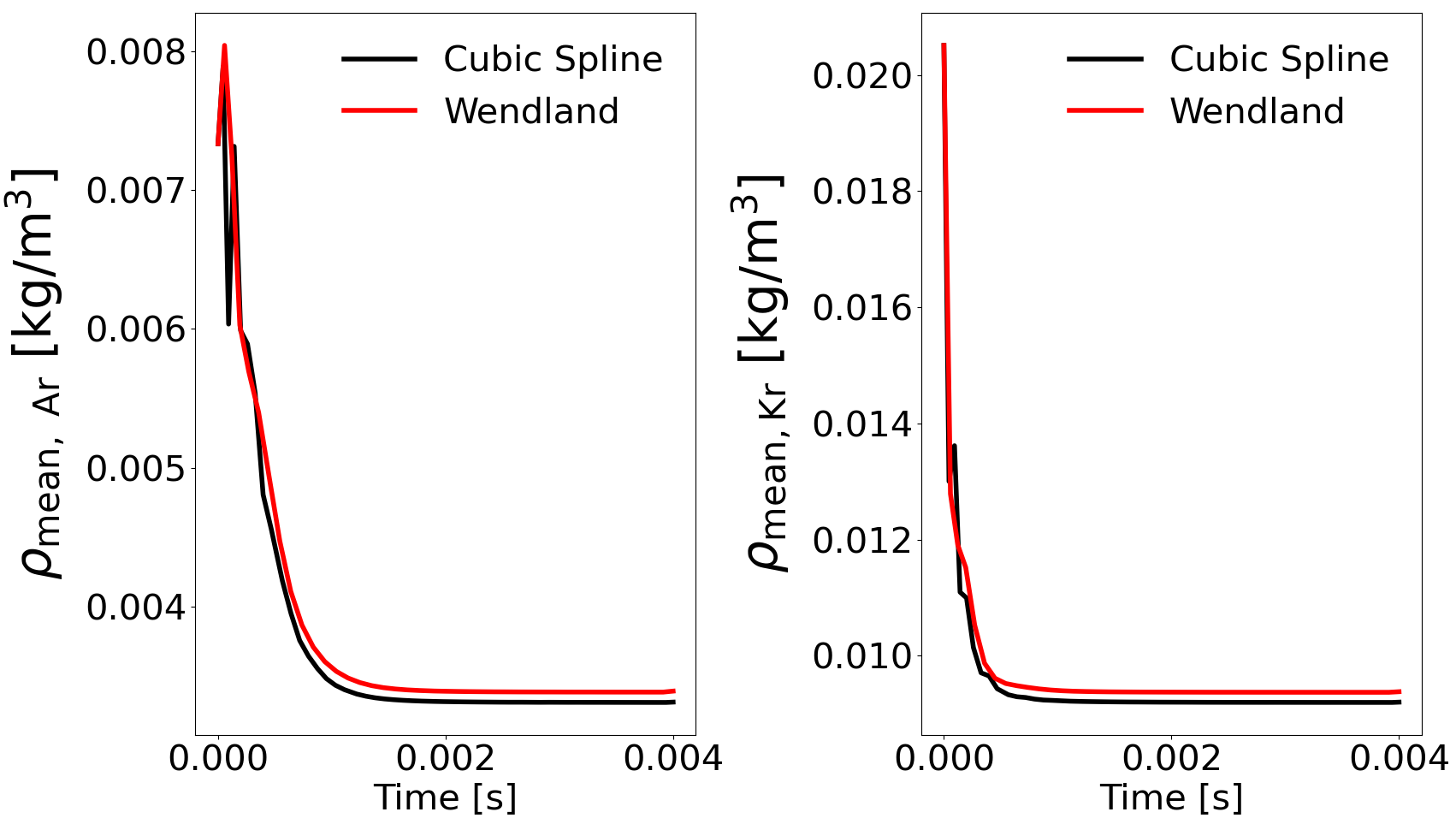}
    \caption{Time evolution of the mean densities of argon (left) and krypton (right) for two different kernel choices: cubic spline (black) and Wendland $C^{6}$ (red). All simulations use the same resolution ($2\times 10^{5}$ particles).}
    \label{fig:kernels}
\end{figure}

To assess the influence of the smoothing kernel on the simulation results, we performed an additional simulation of the same argon–krypton system discussed in Section~\ref{Argon_Krypton}. In this test, the standard cubic spline kernel has been replaced with a Wendland $C^{6}$ kernel \citep{Wendland1995}, while keeping all other parameters identical.
As shown in Fig.~\ref{fig:kernels}, the mean density relaxation value is nearly identical between the two kernels. This suggests that the slight underestimation of the mean densities observed in the numerical tests presented in this work is not related to the kernel choice, but rather to the intrinsic limitations of the geometrical setup adopted in the simulations.

\section{Ideal gas mixing (no collisions)}
\label{sec:Ideal_gas_mix}
\begin{figure}
    \centering
    \includegraphics[width=\linewidth]{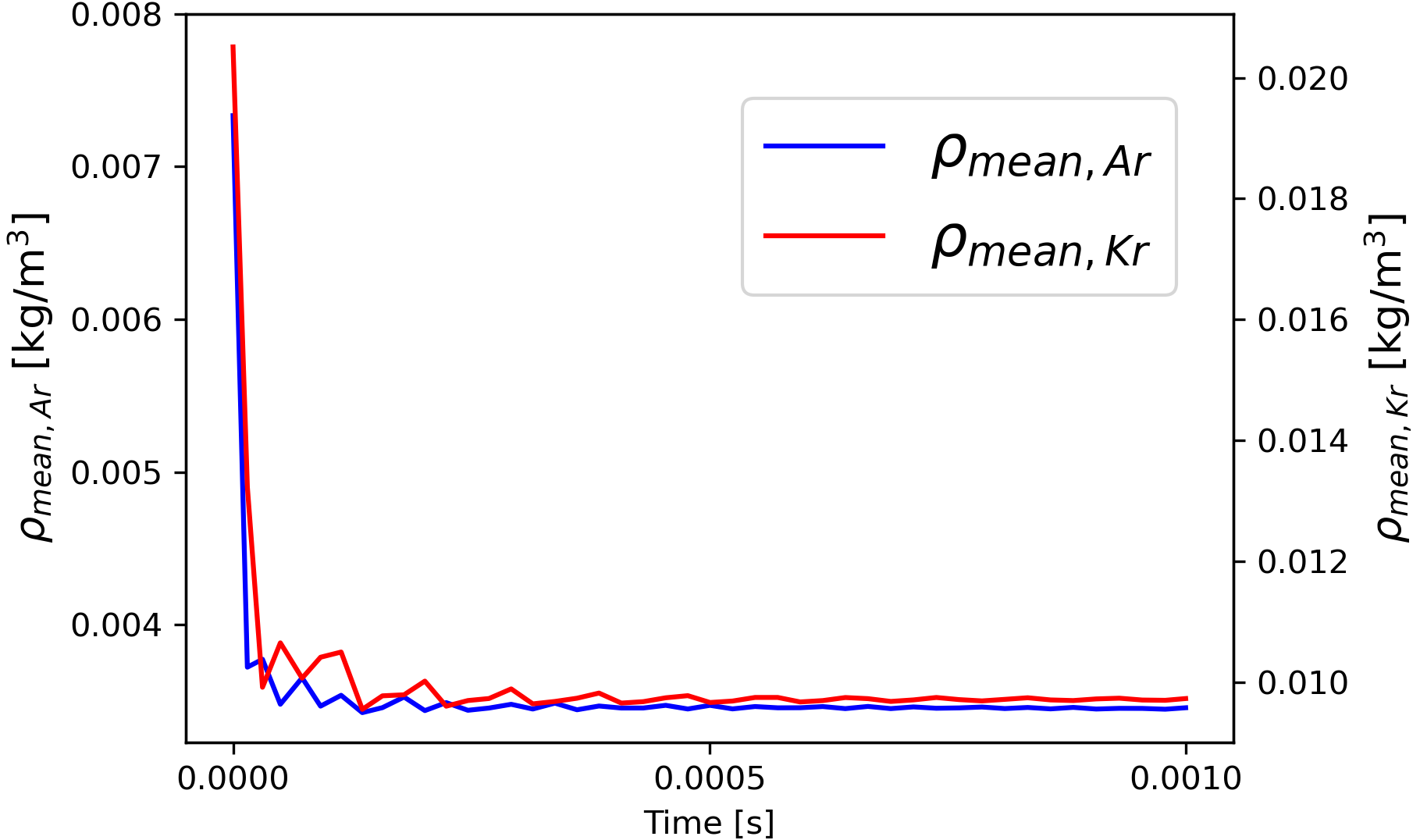}
    \caption{Time evolution of the mean densities of argon (blue) and krypton (red).}
    \label{fig:ideal_gas_densities}
\end{figure}

\begin{figure}
    \centering
    \includegraphics[width=\linewidth]{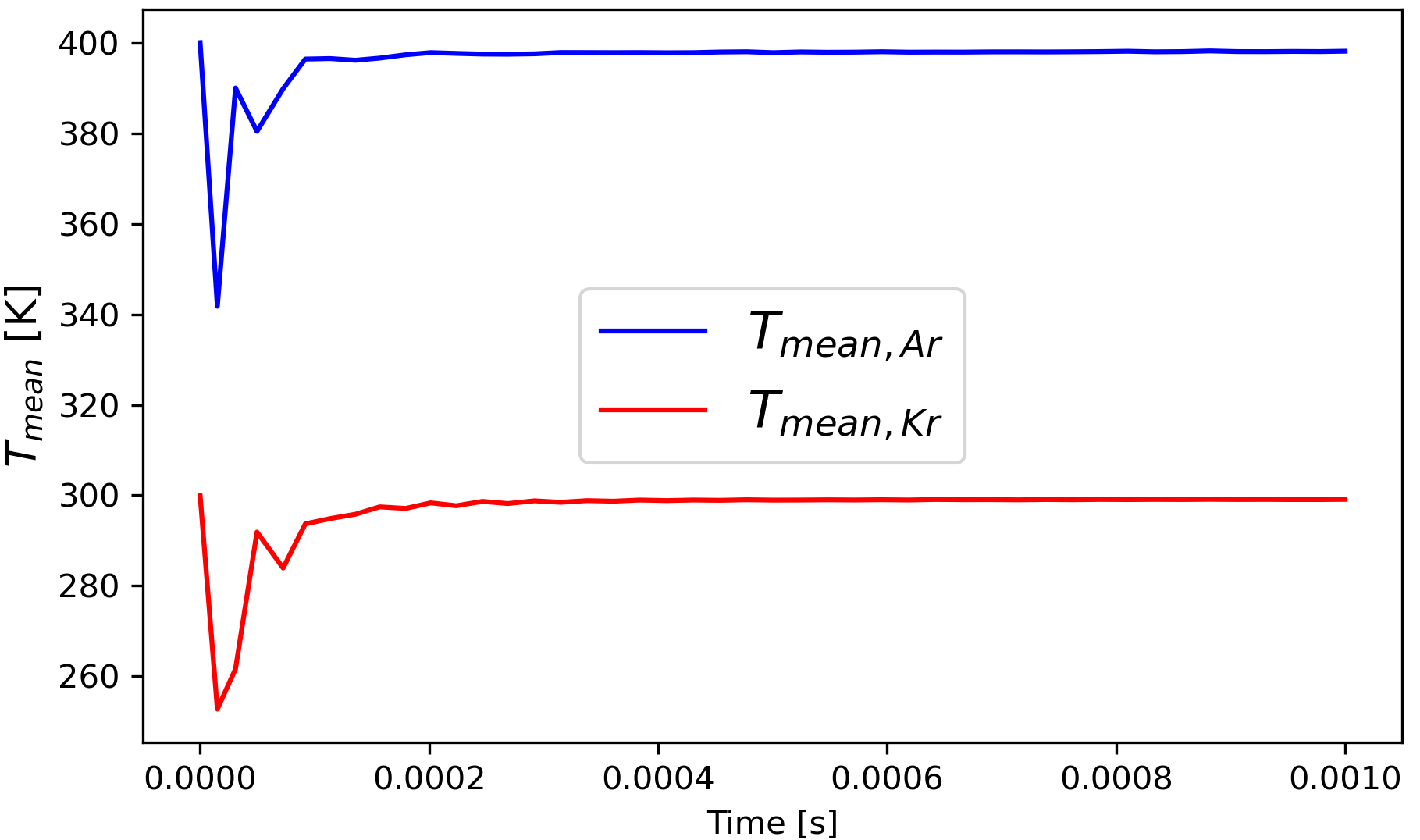}
    \caption{Time evolution of the mean temperatures of argon (blue) and krypton (red).}
    \label{fig:ideal_gas_T}
\end{figure}

Here we report the results obtained from the mixing of two gases ignoring the collisional terms introduced in Eq. \eqref{gas_gas_system} and discussed throughout this paper.\\
The initial setup is the same presented in section \ref{Argon_Krypton}. Nevertheless, looking at Fig.\ref{fig:ideal_gas_densities}, we can observe that the mean gas densities approach the equilibrium value in a very short time, because we are ignoring any collisional effect. Moreover, we do not see any compression of the argon caused by the heavier krypton, because the two gases are behaving as if the other does not exist. For the same reason, the thermokinetic energy of each gas is preserved, leading to an unchanged temperature. The small drop that we see in Fig.\ref{fig:ideal_gas_T} is because at the beginning the velocity of both gases increase due to their expansion, and with them the kinetic energy, but since the total energy of each component is preserved, the internal energy decreased. Finally, when velocities relax to values very close to zero, the internal energy of each gas increases again, reporting the mean temperatures to their initial value.

\section{Entropy evolution and thermodynamic consistency}
\label{sec:entropy}
 In this section, we evaluate the thermodynamic consistency of the model by tracking the evolution of entropy during the mixing process. Since mixing is an irreversible phenomenon, the total entropy of the system is expected to increase over time, in accordance with the second law of thermodynamics.\\
The first law of thermodynamics, expressed in terms of specific (per unit mass) quantities, defines the entropy change as:
\begin{equation}
\mathrm{d}s = \frac{\mathrm{d}\epsilon_{\rm int}}{T} - \frac{P}{T \rho^{2}} \mathrm{d}\rho.
\end{equation}
This relation shows that entropy changes due to both the exchange of internal energy (thermalization) and changes in density (associated with hydrodynamic expansion or compression). Now, from a computational point of view, the total entropy change per unit mass over a timestep is computed as:
\begin{align}
\begin{split}
    \Delta s = &\frac{1}{M_{\alpha}}\sum_{i \in \alpha}m_{i}\left(\frac{\epsilon^{i}_{\mathrm{th, new}} - \epsilon^{i}_{\mathrm{th, old}}}{T^{i}_{\mathrm{mean}}} - \frac{P^{i}_{\mathrm{mean}}}{T^{i}_{\mathrm{mean}} \;\rho_{\mathrm{i,mean}}^{2}}(\rho^{i}_{\mathrm{new}} - \rho^{i}_{\mathrm{old}})\right) + \\
    &\frac{1}{M_{\beta}}\sum_{j \in \beta}m_{j}\left(\frac{\epsilon^{j}_{\mathrm{th, new}} - \epsilon^{j}_{\mathrm{th, old}}}{T^{j}_{\mathrm{mean}}} - \frac{P^{j}_{\mathrm{mean}}}{T^{j}_{\mathrm{mean}} \;\rho_{\mathrm{j,mean}}^{2}}(\rho^{j}_{\mathrm{new}} - \rho^{j}_{\mathrm{old}}) \right).
\end{split}
\end{align}
In this expression, the sum is performed over all SPH particles of gas $\alpha$, indexed by $i$, and of gas $\beta$, indexed by $j$. $m$ denotes the SPH pseudo-particle mass, while $M_{\alpha}$ and $M_{\beta}$ are the total masses of each gas. The differences $\epsilon^{i}_{\mathrm{th, new}} - \epsilon^{i}_{\mathrm{th, old}}$ and $\rho^{i}_{\mathrm{new}} - \rho^{i}_{\mathrm{old}}$ refer to changes in internal energy per unit mass and density between two consecutive timesteps, respectively. Quantities labeled with the subscript "mean" are computed as time-centered averages between successive timesteps.\\
To characterize the entropy evolution, we compute its value at each timestep using the recurrence relation $s(t+dt) = s(t) + \Delta s(t)$, with $s_{0}$ the initial total entropy per unit mass. Since the two gases are initially separated, the total entropy is taken as the sum of the entropies of each component. Assuming ideal monatomic gases, we use the Sackur–Tetrode equation \citep{Emch2002}:
\begin{equation}
    S_{0, i} = k_{\mathrm{B}}N_{i}\; \mathrm{ln}\left[\frac{V_{i}}{N_{i}}\left(\frac{4\pi m_{i}}{3h^{2}}\frac{U_{i}}{N_{i}}\right)^{3/2}\right] + \frac{5}{2}k_{\mathrm{B}}N_{i},
\end{equation}
where $V_{i}$ is the volume occupied by the gas, $N_{i}$ is the number of molecules,  $m_{i}$ the molecular mass, $U_{i}$ the total internal energy, $k_{\mathrm{B}}$ and $h$ the Boltzmann and Planck constants, respectively.
Using this relation and knowing the total mass of the Ar–Kr system discussed in Section \ref{Argon_Krypton}, we estimate an initial entropy per unit mass of $s_{0} \sim 3.15 \times 10^{3} \;\mathrm{J/(Kg \; K)}$. This value is used as a reference for tracking the entropy variation throughout the simulation.
\begin{figure}
    \centering
    \includegraphics[width=\linewidth]{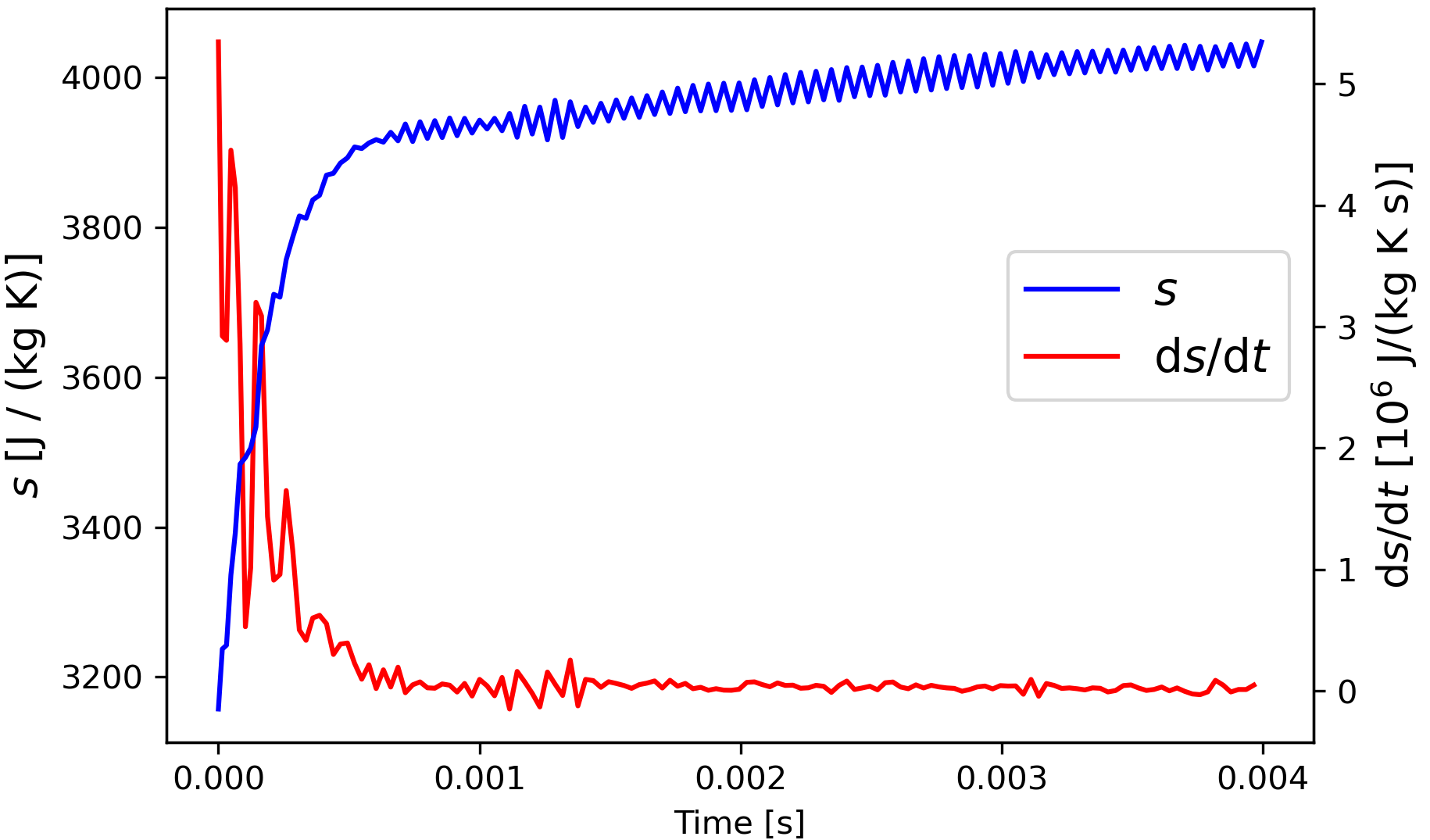}
    \caption{Time evolution of the total entropy per unit mass $s$ (blue line) and its time derivative $\mathrm{d}s/\mathrm{d}t$ (red line) during the mixing of the argon–krypton system described in Section~\ref{Argon_Krypton}.}
    \label{fig:entropy}
\end{figure}\\
Fig.\ref{fig:entropy} shows the time evolution of the total entropy per unit mass, along with its time derivative ($\mathrm{d}s/\mathrm{d}t$).  A continuous increase in entropy is observed, especially during the transient phase, where the entropy production rate remains clearly positive. At later times $(t > 1\times 10^{-3} \; \mathrm{s})$, once thermal and dynamical equilibrium is reached, small fluctuations around zero appear in $\mathrm{d}s/\mathrm{d}t$. These are primarily due to numerical noise and the simple difference-based scheme used to estimate entropy changes. On average, the fluctuations remain slightly positive, leading to a small but persistent increase in entropy. Overall, this behavior confirms that the model respects the second law of thermodynamics, as entropy increases during the irreversible mixing process.
%%%%%%%%%%%%%%%%%%%%%%%%%%%%%%%%%%%%%%%%%%%%%%%%%%

% Don't change these lines
\bsp% typesetting comment
\label{lastpage}
\end{document}